\documentclass[sigconf, table,authorversion]{acmart}
\usepackage[ruled,vlined]{algorithm2e}
\usepackage{amsmath,amssymb}
\usepackage[english]{babel} %localisation
\usepackage{caption,subcaption} %supposedly incompatible with Springer and IOP, IEEETran and ACM SIG
\usepackage{enumitem} %decrease indentation for enumerate/itemize
\usepackage{gensymb}
\usepackage{graphicx}
\PassOptionsToPackage{hyphens}{url}\usepackage{hyperref} %clickable URLS
\usepackage[htt]{hyphenat} %hyphenate \texttt
\usepackage{makecell} %for breaking lines in tables
\usepackage{microtype} %makes text pretty; also condenses
\usepackage{multirow} %multiple rows in tables
\usepackage{placeins} %for making floats not float too much
\usepackage[normalem]{ulem}
\usepackage[redefsymbols=false]{siunitx} %option is for IEEEtran
\usepackage[T1]{fontenc} %ensure proper hyphenation and treatment of math in sentences
\usepackage[utf8]{inputenc}
\usepackage[T1]{fontenc}
\usepackage{arydshln}

\usepackage{tablefootnote}

\newcommand*{\affaddr}[1]{#1} % No op here. Customize it for different styles.
\newcommand*{\affmark}[1][*]{\textsuperscript{#1}}
\usepackage[toc]{appendix}
\usepackage{units}
\usepackage{mathrsfs}   % for using mathscr{F} in the algorithms 
\usepackage{siunitx}
\usepackage{multirow}

\usepackage{footnote}
\makesavenoteenv{tabular}
\makesavenoteenv{table}

\newcommand{\LocOne}{Blacksburg}
\newcommand{\LocTwo}{Arlington}

\newcounter{magicrownumbers}
\newcommand\rownumber{\stepcounter{magicrownumbers}\arabic{magicrownumbers}}

\newcommand{\cut}[1]{}
\copyrightyear{2020}
\acmYear{2020}
\setcopyright{acmcopyright}\acmConference[WiSec '20]{13th ACM Conference on Security and Privacy in Wireless and Mobile Networks}{July 8--10, 2020}{Linz (Virtual Event), Austria}
\acmBooktitle{13th ACM Conference on Security and Privacy in Wireless and Mobile Networks (WiSec '20), July 8--10, 2020, Linz (Virtual Event), Austria}
\acmPrice{15.00}
\acmDOI{10.1145/3395351.3399353}
\acmISBN{978-1-4503-8006-5/20/07}

\begin{document}
\title{Spotr: GPS Spoofing Detection via Device Fingerprinting}

\author{%
Mahsa Foruhandeh, Abdullah Z. Mohammed, Gregor Kildow, Paul Berges, and Ryan Gerdes \\
\affaddr{ Virginia Tech}\\
\affmark[]{\{mfhd,\,abdullahzubair,\,gregor,\,paulberges,\,rgerdes\}@vt.edu}}%

\renewcommand{\shortauthors}{Foruhandeh  et al.}

\begin{abstract}
As the world's predominant navigation system GPS is critical to modern life, finding applications in diverse areas like information security, healthcare, marketing, and power and water grid management.  Unfortunately this diversification has only served to underscore the insecurity of GPS and the critical need to harden this system against manipulation and exploitation.  A wide variety of attacks against GPS have already been documented, both in academia and industry.  Several defenses have been proposed to combat these attacks, but they are ultimately insufficient due to scope, expense, complexity, or robustness.  With this in mind, we present our own solution: \textit{fingerprinting} of GPS satellites.  We assert that it is possible to create signatures, or fingerprints, of the satellites (more specifically their transmissions) that allow one to determine nearly instantly whether a received GPS transmission is authentic or not.  Furthermore, in this paper we demonstrate that this solution detects all known spoofing attacks, that it does so while being fast, cheap, and simpler than previous solutions, and that it is highly robust with respect to environmental factors. 
\end{abstract}


% <concept_id>10002978.10002997.10003000.10011611</concept_id>
% <concept_desc>Security and privacy~Spoofing attacks</concept_desc>
% <concept_id>10002978.10003014.10003017</concept_id>
% <concept_desc>Security and privacy~Mobile and wireless security</concept_desc>

% \ccsdesc[500]{Security and privacy~Spoofing attacks}
% \ccsdesc[500]{Security and privacy~Mobile and wireless security}


\maketitle

\section{Introduction}

GPS, in its original role as a navigational aid, can be found in everything from smartphones to unmanned aerial vehicles (UAVs).  This alone would merit the attention of security researchers.  Today trackers use it to trace the movement of entities, games and streaming services use it for localizing advertisements, secure facilities use it to ``geo-fence'' their equipment, and power grids use it for phase synchronization.  Put simply, there are very few aspects of modern life that are not deeply dependent upon GPS.

In order to spur the growth of GPS and ease its adoption by the private sector, all aspects of the design and implementation of the so-called L1 C/A (``Coarse Acquisition'') signal were made public.  GPS---``...the most popular unauthenticated protocol in the world.''\cite{humphreys2012statement}---has since become the primary Global Navigation Satellite System (GNSS).  Allowing anyone to acquire a detailed understanding of GPS with very little effort, combined with the relatively trusting nature of GPS transmitter/receiver interaction, makes the system ripe for manipulation and catastrophic exploitation.  As such, we must harden GPS as a matter of national security.

Several methods that attempt to provide post hoc security for GPS exist in the literature \cite{psiaki2016gnss}. Our approach uses  physical-layer identification (PLI), aka \textit{device fingerprinting}, and falls within the category of correlation profile anomaly detection. There is a long history in using device fingerprinting for securing transceivers \cite{gerdes2006device} and satellites are an example of such transceivers. We introduce Spotr, a GP\textbf{S} s\textbf{p}o\textbf{o}fing detec\textbf{t}ion via device fingerp\textbf{r}inting that is able to determine the authenticity of GPS signals based on their physical-layer similarity to signals that are known to have originated from GPS satellites. We extract strong features from the outputs of the complex correlators common to all GPS receivers and use them to generate templates for the genuine satellite signals, which we call \textit{fingerprints}. Finding a strong fingerprint for genuine GPS signals is non-trivial, since a spoofing adversary always tries to mimic the characteristics of the authentic signal to the best of their capability. We introduce simple (but difficult to spoof) features here, yet we are able to detect the most powerful spoofing attacks with high accuracy and in a timely fashion. Our main contributions are: 
\begin{itemize}

    \item  We detect all the existing spoofing attacks on different dimension of GPS receivers in literature (time and/or position) using a single channel/receiver.
    \item  We are able to track a genuine satellite over different days, environmental conditions, and locations, such as urban and rural settings, with and without multipath propagation effects. Our method does not require channel modelling, which reduces its complexity. We are only limited to having a good observation of the signal to perform device fingerprinting. 
    \item  Since we use fingerprinting, we do not impose a lower-bound on the range of spoofing detection. This is an improvement to the state-of-the-art spoofing detectors which fail to find attacks up to the range of \unit[1000]{m}. 
    
\end{itemize}
   Our spoofing detector functions properly regardless of the number of devices that the spoofer is utilizing. This enables us to detect coordinated attacks on multilateration systems (estimation of distance based on time of arrival of waveforms travelling at a known speed \cite{2017UnderstandingApplications}). Our approach is purely passive with no modifications required to the existing GPS protocol, satellite orbits or uplink/downlink communication channels. We also make our dataset of genuine and spoofed GPS signals available for the community. %to employ for further study.
\subsection{Related Work}
\label{Sec::relatedwork}
The public availability of civil GPS implementation details has made it trivial to accurately and reliably (re)create navigational data intelligible to any GPS receiver.  What's more, these receivers are not only inherently trusting and accept any GPS signal they can demodulate, but the automatic gain control (AGC) system in the standard GPS receiver is designed to favor (lock onto) the strongest signals. While this behavior may seem intuitive, it also assumes -- dangerously -- a benign environment which means that a malicious attacker can easily assert control over a receiver simply by using a closer, more energetic signal to overpower the weaker, authentic signals.  This is known as the \textit{overpower attack} and it is the focus of most detection and mitigation strategies.  Numerous lines of defense are given in \cite{psiaki2016gnss,humphreys2012statement} which can be broadly categorized as cryptographic, hardware-specific or signal processing-based solutions. 

Non-predictable modulation techniques are proposed to overcome the problem of open signal structure in \cite{pozzobon2011keeping,wesson2012practical}. These are mostly cryptography-based methods that, while valid, depend on modifications to the space segment.  As such, the costs involved with redevelopment or even modification of satellites in situ render these solutions impractical despite their effectiveness. 

Jamming, often seen as a threat to GPS, can also be a standalone, non-cryptographic line of defense against GPS spoofing. One such approach uses jamming-to-noise (J/N) sensors that are inexpensive and easy to build. The sensors mark activity as malicious if the energy of the in-band signal exceeds a given threshold, which forces the spoofer to limit its power and makes the overpower attack far more difficult to maintain.  This defense is probabalistic, however, as a carefully crafted spoofer could still employ more nuanced power adjustments and it would be unwise for one to assume anything less than maximum determination in an adversary \cite{wang2019gnss, humphreys2012statement}. 

Hardware-based solutions include those that augment GPS receivers with additional antennae or inertial sensors. The single/ multiple antenna methods are fast and reliable, but are expensive and require modifications to current receivers and redesign of future receivers \cite{montgomery2009multi,jansen2016multi,tippenhauer2011requirements}. For their own part, the inertial sensors create a new accuracy concern as their measurements are temperature sensitive and thus subject to drift depending on environmental factors \cite{el2007temperature}. 

Another set of defense methods implement spread spectrum security code (SSSC) or navigation message authentication (NMA) on wide area augmentation system (WAAS) \cite{enge1996wide}. WAAS is an air navigation aid developed by the Federal Aviation Administration (FAA) to augment GPS in order to increase its accuracy, integrity, and availability. With no modifications to the space segment of the GPS, this is strong enough to stop the spoofer because SSSC is a strong high rate security code, however, it cannot authenticate a full three-dimensional navigation solution. Long delay is another drawback of SSSC if used in aviation. NMAs, easier to implement compared to SSSC, are slightly less secure with equal delay  \cite{kroner2010hardening}. 

A recent alternative solution is the Multi-System Multi-Frequency defense. Secondary to the U.S. Air Forces's ongoing GPS modernization project, civilian GPS signals are now being transmitted on other bands such as L2 and L5 in addition to the legacy L1. The signals at L2 and L5 can be used for consistency check of the signal at L1, which makes spoofing more challenging \cite{montenbruck2017multi}.% The main drawback, however, is that the multi-frequency receivers available on the market at this time can be quite expensive . 

Correlation profile anomaly detection is a different approach which mainly relies on the difficulty of suppressing the genuine GPS signals and the fact that even the strongest spoofer fails at exactly mimicking the authentic GPS signal's behaviour. This helps to find anomalies and use them to detect spoofing activities. It is a low cost software solution which eliminates the need for additional hardware. This method is known to function efficiently for stationary receivers, while its performance is influenced by propagation effects of the wireless channel such as multipath and fading \cite{humphreys2012texas}. In a similar way, a PLI-based method for spoofing attacks is given at \cite{moser2016investigation} where the frequency offset and transient phase noise of the attacker’s radio are the main features for detection, which are weaker than the correlator outputs used in the present work as they are more susceptible to spoofing.

Some techniques attempt to secure the GPS C/A signals using the existing military signals which are not cost-efficient \cite{psiaki2013gps}.

Numerous attacks have already been demonstrated on the GPS system, strongest of which is a seamless lock takeover attack \cite{tippenhauer2011requirements} where the attacker takes over the target receiver gradually, by avoiding abrupt changes which might lead to detection. SPREE is introduced at \cite{ranganathan2016spree} which is a spoofing resistant GPS receiver, able to limit the state-of-the-art spoofing attacks up until 2016 including seamless lock takeover attack \cite{tippenhauer2011requirements} . It introduces a spoofing detection technique named auxiliary peak tracking (APT) which operates next to the second module of SPREE named navigation message inspector (NMI). Unlike ordinary, SPREE acquires and tracks all of the signals (the strongest and the weaker ones), and uses NMI to look for discrepancies in the content of the navigation messages to detect possible attacks. It contributes to the security of GPS systems by limiting the range of spoofing attacks on position to the radius of \unit[1000]{m}. This is accomplished at the cost of using more than one channel to acquire, track and decode each satellite's signal, which also requires modifications to the GPS receiver. The main challenge for SPREE is to distinguish the auxiliary peak of a spoofing signal from that of a multipath component of a genuine GPS signal. This makes it difficult for SPREE to cancel the spoofing signal due to uncertainties at identifying the source of the signal.
SPREE also relies on the presence of the authentic signals (assuming that these signals are not already suppressed by the attacker) which makes it vulnerable to attacks which nullify the authentic signals.

% VSD ---------------------------------------------------------------------------
In a similar way, Vestigial Signal Defense (VSD) \cite{wesson2011evaluation} attempts to detect spoofing based on analyzing the distortions present in the output of the receiver's complex correlation function. VSD faces challenges to distinguish these distortions from legitimate multipath components. Other similar detection strategies which rely on inherent spatial characteristics of the received signal ( e.g. angle of arrival \cite{montgomery2011receiver}) face the same challenges. Simpler spoofing detection strategies that look for anomalies in the physical-layer characteristics of the received signal such as abrupt changes in power level of the received signal or the AGC value \cite{akos2012s} are usually not able to detect a spoofer which has proper control over its signal.

\begin{figure*}[t]
\centering
\subcaptionbox{\label{fig::overview}}{\includegraphics[width=0.375\textwidth]{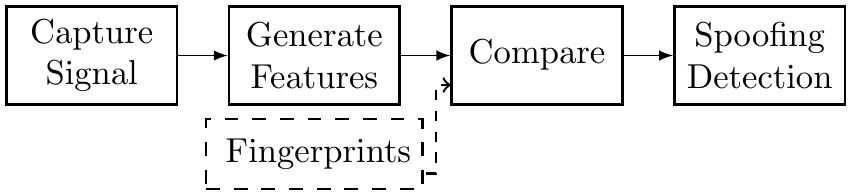}}
\hspace{5pt}
\subcaptionbox{\label{fig::receiver}}{\includegraphics[width=0.4\textwidth]{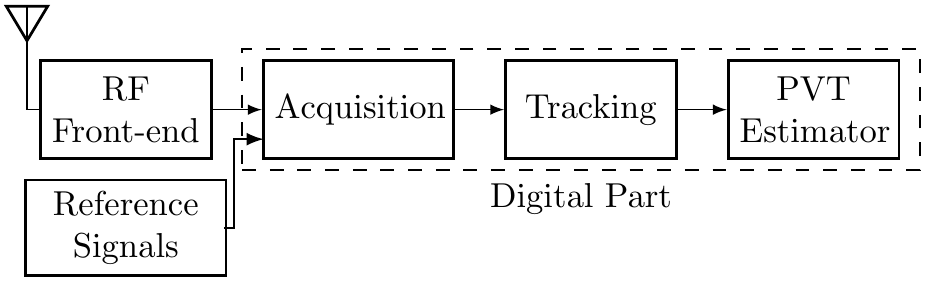}}
%\hspace{-5pt}
%
\subcaptionbox{\label{fig::rf}}{\includegraphics[width=0.43\textwidth]{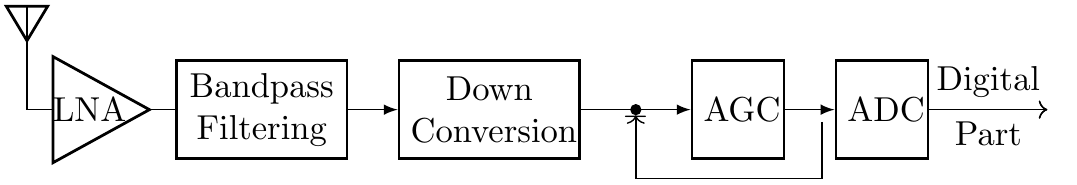}}
\subcaptionbox{\label{fig::agc}}{\includegraphics[width=0.55\textwidth]{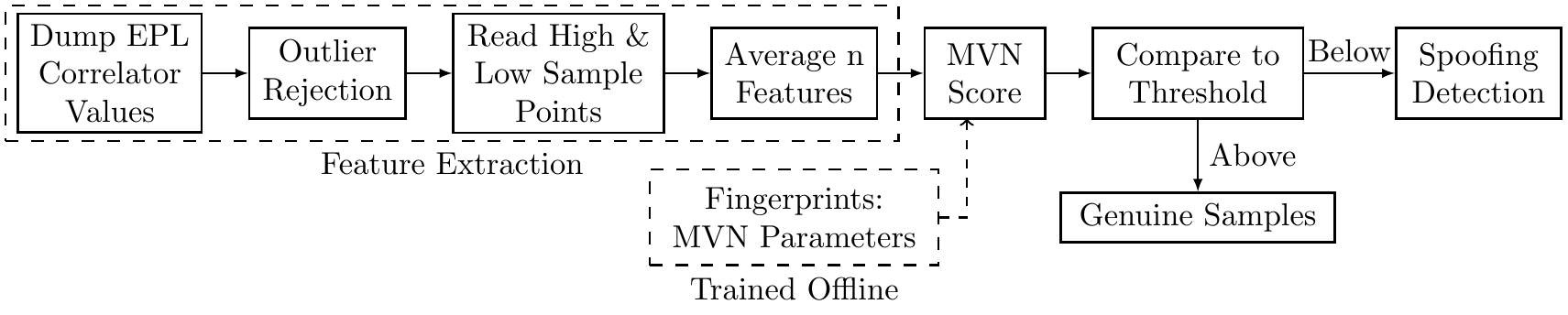}}
\caption{(a) Overview of satellite fingerprinting, spoofing detection. (b) Block diagram of GPS receiver. (c) RF front-end of GPS receiver. (d) Spotr overview of defense.}
\end{figure*}

\subsection{Paper Structure}
 Section \ref{sec::attaCk} covers the types of attacks that we address. Sec. \ref{sec::Spotr} provides a background on GPS and describes our defense methodology where we introduce Spotr with device fingerprinting for detecting spoofing attacks. Same section covers the feature extraction and the real-time spoofing detection process. Data collection is explained at Sec. \ref{sec::data}. Experimental validation of Spotr is discussed at Sec. \ref{sec::rslt}, where we also do a feature stability analysis to examine their consistency on different conditions before we conclude at Sec. \ref{sec::conc}.

\section{Attack Model}\label{sec::attaCk}
We envision a physical-layer attack where the objective of the attacker is to induce a different position, velocity, and/or time (PVT) solution in a targeted civilian GPS receiver. A spoofing attacker is able to craft digitally valid GPS signals indistinguishable from  genuine ones, inject them into a wireless channel, and produce a valid PVT at the victim. A replay attacker, on the other hand, captures analog data from genuine satellites and replays it to the victim. The latter is a stronger attack since it replicates the subtle intrinsic features of a genuine signal. The attacker is aware of the location of the target receiver. In both cases, if a receiver locks to the fake satellite (signal) then the attacker can choose a desired/altered PVT. We note that GPS signal generators that would enable this are readily available.  We do not place any artificial restrictions on the attacker's ability to generate and inject signals: the spoofed signal may be greater in power than the genuine GPS signals (``over-powered''), at a slightly higher power level (``matched-powered''), at a lower power level  (``under-powered'') in case of a near-instantaneous switch from an exclusively authentic stream into an exclusively spoofed one, and/or phase aligned to enable ``seamless'' takeovers \cite{humphreys2012texas,humphreys2016texbat}. This maximizes the chances of an attacker to remain undetected while enabling all of the attacks discussed in \cite{ranganathan2016spree}. Finally, a replay attacker is able to adjust power levels and create signals with correlation outputs as close as possible to genuine ones.

\section{satellite spoofing detection} \label{sec::Spotr}
Here, we will describe our physical-layer based spoofing detection scheme, Spotr. Feature extraction is the most essential step in designing a PLI system. We start this out by finding simple, yet strong features that make the detection possible while cutting down on the complexity. Next, we use an offline training-testing scheme to generate thresholds and use them for real-time detection of spoofed or tracking of genuine GPS signals. Fig.~\ref{fig::overview} shows overview of a PLI system in context of satellite spoofing detection, detailed below.

\subsection{GPS Receiver}\label{sec::digreceiver}
Fig.~\ref{fig::receiver} illustrates a block diagram of a GPS receiver. The main components of GPS receiver are (i) RF front-end, (ii) acquisition module, (iii) tracking module, and (iv) PVT estimator module. The RF front-end investigates if satellite signals are available. The acquisition module searches for the delayed versions of any satellite's pseudo random noise (PRN) sequence while estimating the Doppler shift and compensating for it. PRNs are codes used to differentiate the signals generated from different satellites in a code division multiple access (CDMA) system. The role of a tracking block is to follow the evolution of the signal synchronization parameters: code phase, Doppler shift and carrier phase. As a main component of tracking block, the VOLK-Library \cite{Fernandez-PradesGNSS-SDR} from gnss-sdr, is responsible for running the delay locked loop (DLL) and phase locked loop (PLL) at the GPS receiver. It runs at a varying speed in a feedback system based on the quality of tracking in that instance. The PVT estimator uses this to report a solution on position, time and velocity. The most common method to do so is called multilateration, where an accurate estimate of location coordinates of a GPS transmitter plus time requires data from at least four satellites \cite{keysightGNSSapplicationnote}.  % , and at the same time could be used as a foothold by the attackers

\subsection{Feature Selection Rationale}\label{sec::observpoint}
GNSS receivers are composed of an analog RF front-end and a digital part. Fig.~\ref{fig::rf} illustrates the block diagram of the RF front-end with an automatic gain control (AGC) component right before the Analog-to-Digital converter (ADC). AGC is an adaptive feedback loop system that uses a variable-gain amplifier (VGA) in-order to provide consistent inputs to the ADC \cite{bastide2003automatic}. This consistency makes correlators a good observation point for feature extraction. A potential optimal detector, to determine the presence of a valid signal, is a matched-filter, which maximizes the signal-to-noise ratio of a known input signal in additive white Gaussian noise (AWGN) \cite{couch1997digital}.

\begin{figure*}[t]
	\centering
	\subcaptionbox{\label{fig::GPS_receiver_EPL}}{\includegraphics[width=0.6\textwidth]{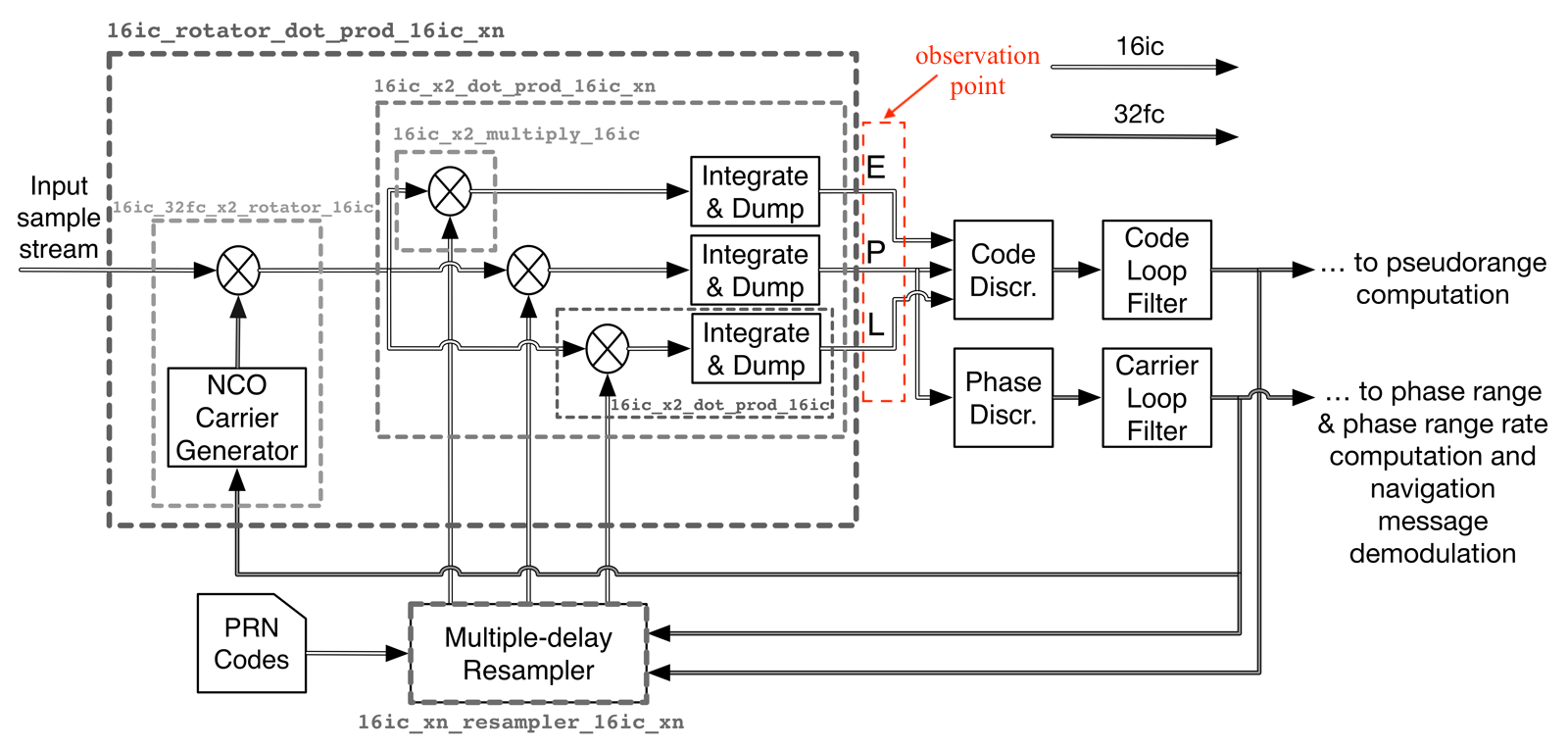}} 
	\subcaptionbox{\label{fig::EPL_samples}}{\includegraphics[height=5cm,width=0.35\textwidth]{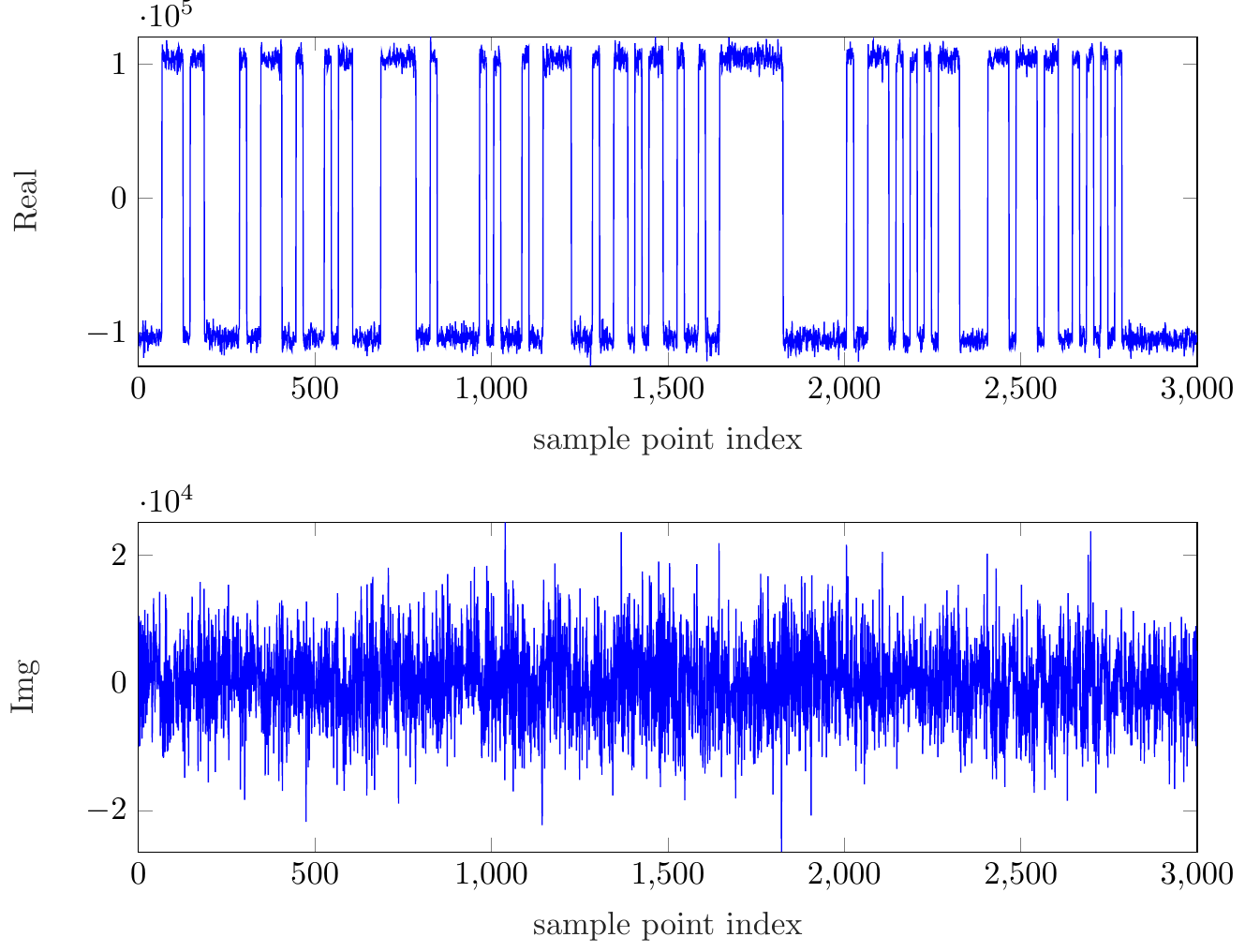}}
	\caption{(a) EPL correlator outputs for GPS digital receiver \cite{Fernandez-PradesGNSS-SDR}, the observation point of Spotr. (b) Output of P correlator.}
\end{figure*}

 Earlier work in fingerprinting satellites uses weaker features, such as modulation imperfections or AGC parameters, that are known to be vulnerable to low-cost spoofing attacks \cite{danev2010attacks,broumandan2015spoofing}.  Correlator outputs \footnote{GPS uses direct sequence spread spectrum (DSSS) modulation. For the purpose of demodulation and de-spreading, correlators are required in the receiver design. Therefore, they are an inherent part of any GPS receiver.}, however, require better resourced attackers with arbitrary waveform generators (AWGs) \cite{gerdes2012physical}. In an effort to mimic such an attacker we perform matched-powered replay attacks in Sec.\ \ref{eval::adept}. Given their security, together with the theory of operation of the AGC, make the correlation outputs reliable features (Fig. \ref{fig::GPS_receiver_EPL}).

\subsection{Feature Extraction}
\label{sec::ftrs}
Based on documentation of GPS, we have samples after I-Q demodulation at the RF front-end which contain information on the P(Y) and C/A code, the military and civilian codes, respectively. The output of the matched-filter correlation, $\mathrm{M}_{\mathrm{o}}$ would be % a conjugate multiplication given by
\begin{equation}
    \label{eq::conjmul}
    \begin{split}
    \mathrm{M}_{\mathrm{o}} &= (I+\mathrm{j}\,Q)\,\times (I_{\mathrm{rep}}-\mathrm{j}\,Q_{\mathrm{rep}}) \\
    & = (I\times I_{\mathrm{rep}}+Q\,\times Q_{\mathrm{rep}})+\mathrm{j}\, (Q\times I_{\mathrm{rep}}-Q_{\mathrm{rep}}\times I) 
    \end{split}
\end{equation}
where rep shows the replica of the PRN code in the receiver. Given the right-hand-side of the equation, if there is a match, the real part will hold a large enough value and the imaginary part will be close to zero which will result in a lock at the tracking block. There are three of such correlators shown in Fig. \ref{fig::GPS_receiver_EPL}, named early, prompt, and late correlators called E, P and L. A sample of the real and imaginary part of the P correlator is given at Fig. \ref{fig::EPL_samples} where we will focus on the real part which holds most of the information from the satellite. The ideal value of the imaginary part is zero \cite{misra2006global}. 

\begin{figure*}[t]
\centering
\subcaptionbox{\label{fig::clt_cn0}}{\includegraphics[height=4.75cm,width=0.33\textwidth]{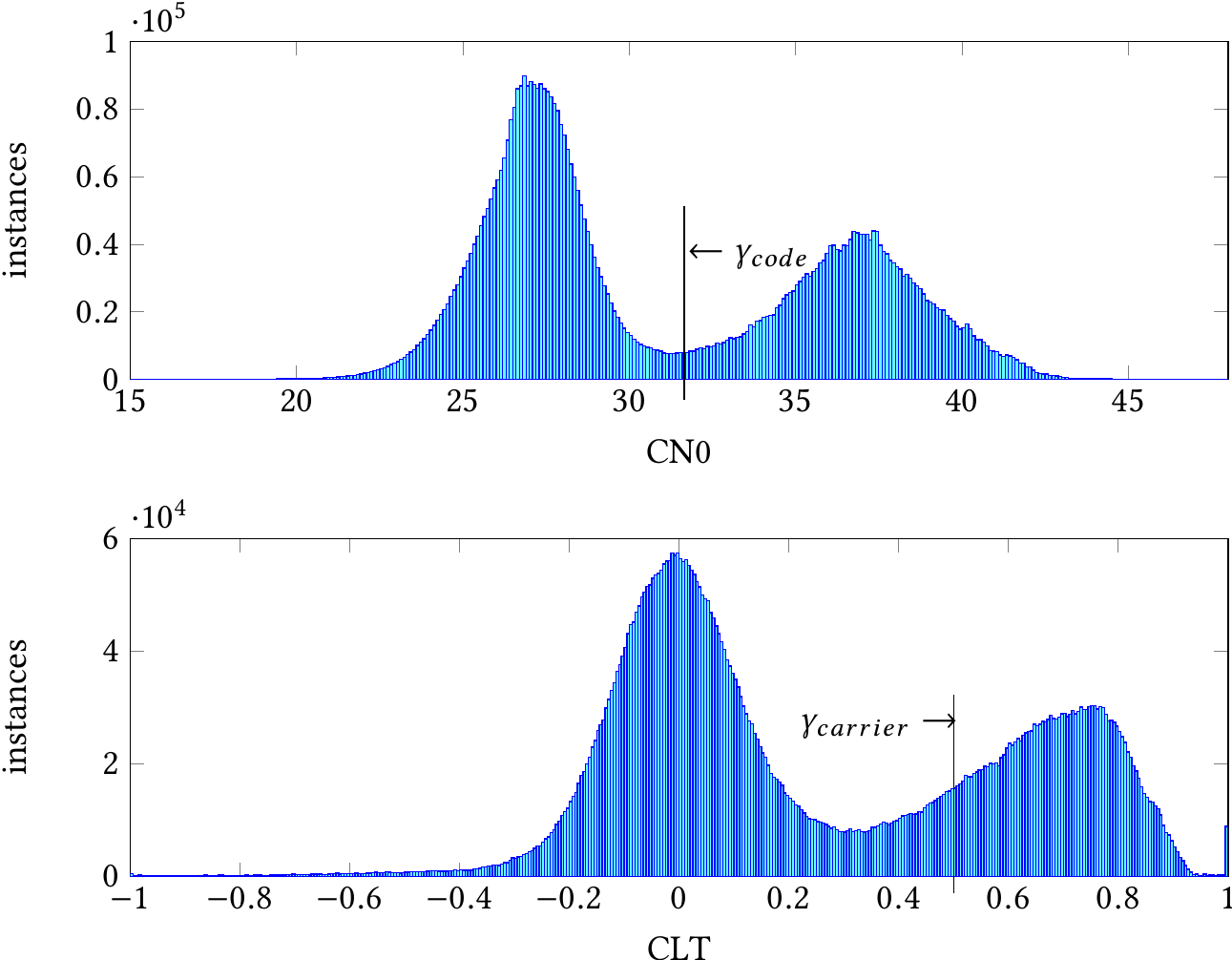}}
\subcaptionbox{\label{fig::L_hist}}{\includegraphics[width=0.33\textwidth]{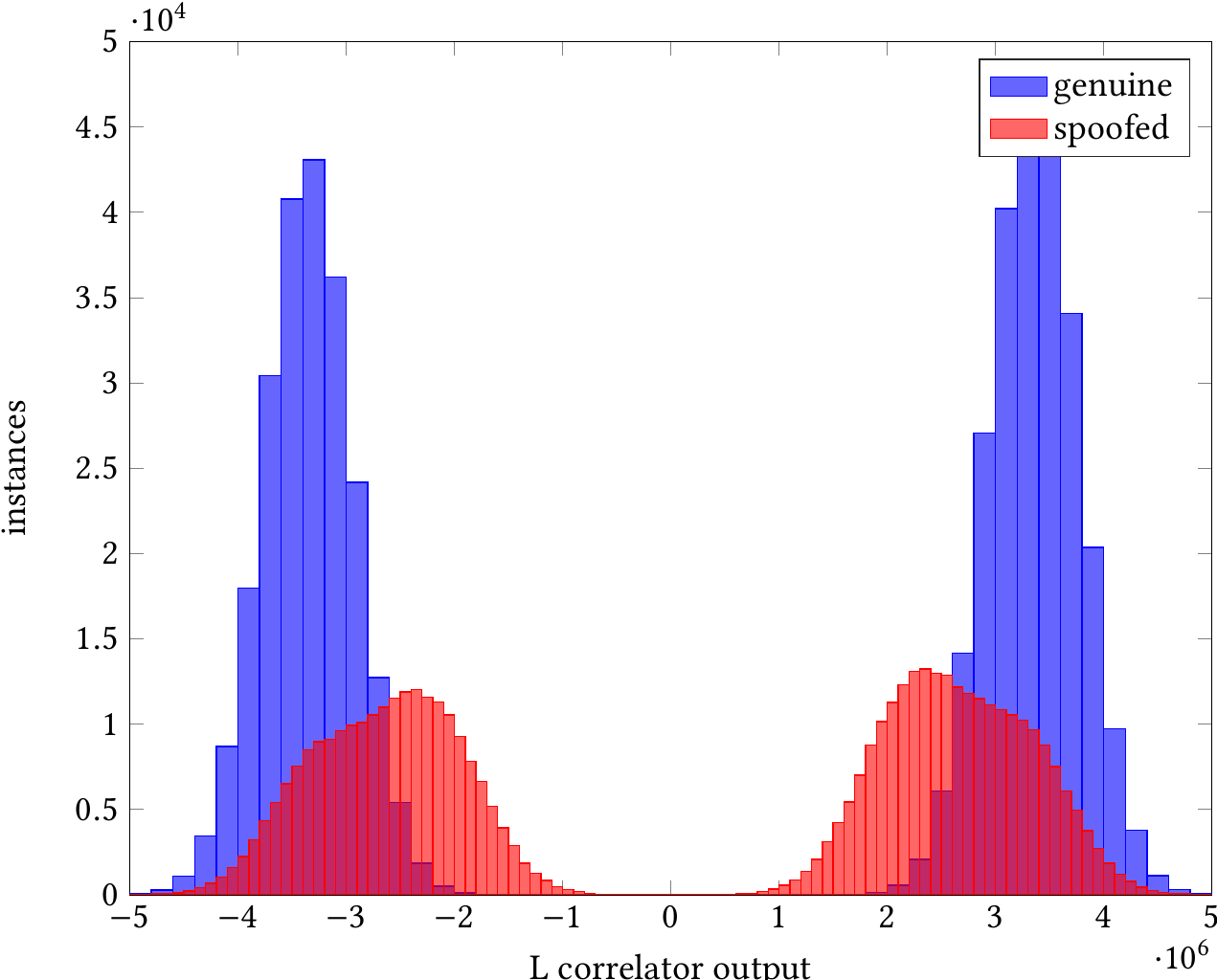}}
\subcaptionbox{\label{fig::multi_sp}}{\includegraphics[height=4.75cm,width=0.33\textwidth]{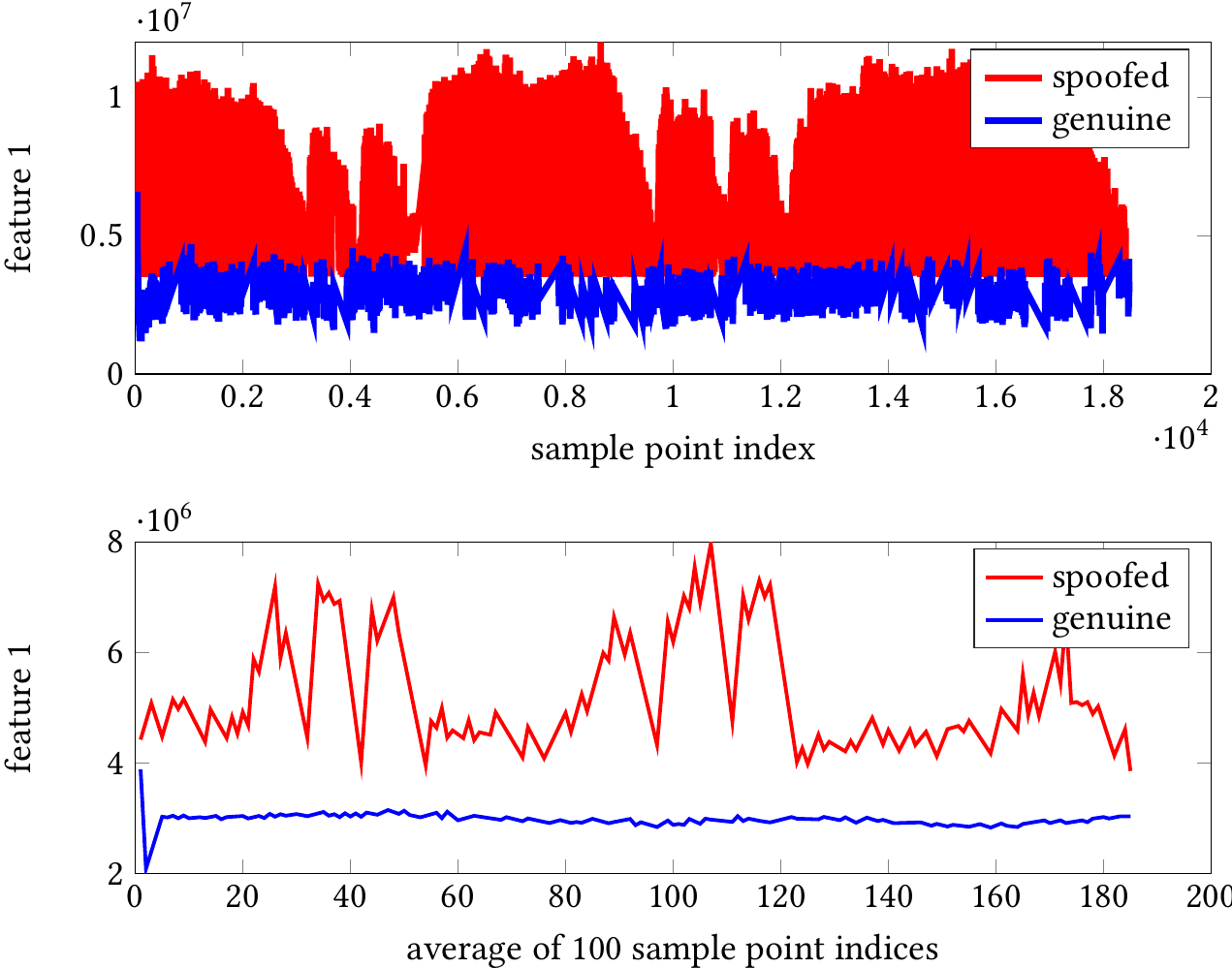}}
\caption{(a) Histogram of CN0 and CLT for SatGrid:S10 (see Tables \ref{table::datasets1} and \ref{table::datasets2} for details on the datasets). (b) Late correlator output for PRN23 of TexBat:S8. (c) Using average of multiple features to create separation between spoofed and genuine signals.}
\end{figure*}

The gnss-sdr receiver uses two metrics to validate the signal quality to generate a lock, named the code-lock-detector (CN0) and carrier-lock-detector (CLT). The CN0 test is defined as 
\begin{equation}
    \label{eq::cn0}
    \widehat{C/N}_{0_{dB-Hz}} \overset{\text{lock}}{\underset{\text{no lock}}{\gtrless}} \gamma_{code},
\end{equation}
where $\widehat{C/N}_{0}$ is an estimate of $C/N_0 =\frac{C}{\frac{N}{BW}}$ ($C$ is the carrier power, $N$ is the noise power and $BW$ is the bandwidth of observation) which is the carrier-to-noise density ratio and refers to the ratio of the carrier power and the noise power per unit of bandwidth. The threshold $\gamma_{code}$ is set to 32 dB-Hz here. The the lock detector test for the carrier tracking loop is defined as, 
\begin{equation}
 \label{eq::gamma}
\cos(2\widehat{\Delta \phi}) \overset{\text{lock}}{\underset{\text{no lock}}{\gtrless}} \gamma_{carrier}, 
\end{equation}
where $\Delta \phi = \phi - \hat{\phi}$ is the carrier phase error. If the estimate of the cosine of twice the carrier phase error is above a certain threshold, the loop is declared in lock. The threshold $\gamma_{carrier}$ is set to 0.5 radians. This is referred to as CLT in our data collection software at Sec. \ref{sec::data}. Fig. \ref{fig::clt_cn0} shows the histograms of the CN0 and CLT, which are used for deciding on the thresholds. 

After filtering out the sample points that fail the above lock tests, we look into the real values of the EPL correlator outputs to generate strong features for each satellite. For each correlator, we separate the points above zero and the ones below zero as high/low sample points and we will have features of six dimensions which is high and low sample points of EPL correlator outputs. Feature one is high of E correlator, feature two is low of E correlator, feature three is high of P correlator, feature four is low of P correlator, feature five is high of L correlator, and feature six is low of L correlator. Histogram of features 5 and 6 is shown at Fig.~\ref{fig::L_hist}. %for both genuine and spoofed signals. 

\begin{algorithm}[b]
	\caption{Feature Extraction}\label{FtrAlg}
	Generating a feature template \textbf{\textit{F}} for each satellite\\
	\For {\text{$\textit{Each PRN }\, i>1 $} }{
		Collect the outputs of EPL correlators\\
		%\textbf{goto} \text{upper/lower level sample points Alg. \ref{HighLow}.}\\
		$F_{i} \gets \textit{high and low of EPL outputs} $\\
		Apply code-lock-detector test (Eq.\ref{eq::cn0})\\
		Apply carrier-lock-detector test (Eq.\ref{eq::gamma})\\
	}
	{$F \gets F \cup {F_{i}}$}
\end{algorithm}
\subsection{Multivariate Normal Distribution}

As discussed in Sec. \ref{sec::ftrs} we are dealing with high dimensional data. In order to describe the characteristics of this data we apply a multivariate normal distribution (MVN) \cite{bishop2006pattern} to each dataset. For a set of $\{\mu,\Sigma\}$ that we assign, the MVN score is calculated by Eq.~\ref{eq::mvn}. Gaussian distribution fits the closest to our features, one example of which is plotted in Fig.~\ref{fig::L_hist}. We hypothesize that the features are correlated and we intend to capture this relation between them, that is the reason we choose MVN as the scoring metric. It is the main metric for measuring similarity in our work and translates to how close each observation from a dataset is to a specific distribution. 

\begin{equation}
\label{eq::mvn}
    f_{x}\left ( x_{1}, x_{2}, \cdots ,x_{k} \right )= \frac{\mathrm{exp}\left ( -\frac{1}{2}   \left( x-\mu\right)^{T}  \Sigma^{-1}   \left( x-\mu\right)        \right)}{\sqrt{\left( 2\, \pi\right )^{k}\left | \Sigma \right |}}.
\end{equation}

%\subsubsection{Using Multiple Sample Points}
\label{sec::multi_sp}
To guarantee a clear separation between the genuine GPS signals and the spoofed ones in different weather conditions or locations, we use the average MVN score from multiple sample points. Fig.~\ref{fig::multi_sp} shows the effect of using a single sample point for comparison versus using the average of 100 sample points. The lower graph shows a better separation between the genuine and spoofed signals. Hence, in the experimental evaluations at Sec. \ref{sec::rslt} we will use multiple sample points (shown by \textit{n}) and average their MVN scores.%An alternative approach could be using summation of MVN scores for multiple sample points. This adds the same complexity as that of the former, hence, we proceed with the former approach (i.e. average of multiple sample points). 

\subsection{Real-time Detection}
We design our PLI-based spoofing detector in two main steps. First, we do a training and testing procedure which results in thresholds for identification of genuine satellite signals from the spoofed ones. Next, we train the detector to trigger once an anomaly is observed using these thresholds in real-time manner.  

\textbf{Training:} In this step we use 400 sample points observed from a spoof-free dataset and generate a template set of features calculated according to Sec.  \ref{sec::ftrs}. We use these training features to fit an MVN distribution for the genuine GPS signals. We assume training is secure, which is accomplished by collecting data at known locations.

\textbf{Testing:} We calculate the MVN score for each observation from the spoofed GPS dataset. A low MVN score is an indication of spoofing in the GPS signal. To define a threshold and finalize the spoofing detection process, we conduct a binary search algorithm to find a threshold that corresponds to equal error rates (EER). EER is a measure of performance for bio-metric systems which indicates a condition in which the false positive rates (FPR) equal to false negative rates (FNR). An ideal value for EER would be zero \cite{bolle2013guide}. 
\begin{algorithm}[b]
	\caption{Overview of Proposed Approach: Training, Testing, and Real-time Detection/Tracking}\label{alg::detection}
	    \textbf{Training:}\\
	\For {$\textit{dataset i }$}{
		{$F_{i}  \gets$  feature extraction Alg. \ref{FtrAlg}}\\
		{$\mathfrak{F} \gets \mathfrak{F} \cup {F_{i}}$  trained }
	}
        {$\mathrm{MVN}(\mu, \Sigma) \gets$ fit MVN on $x\%$ of $\mathfrak{F}$} \\
        {$\mathrm{sc_{MVN}} \gets$ the density values for the remaining $\mathfrak{F}$ on $\mathrm{MVN}(\mu, \Sigma)$ }\\
        {$threshold \gets$ EER analysis on $\mathrm{sc_{MVN}}$ } 

	\setcounter{AlgoLine}{0}
	
	\textbf{Testing:}\\
	{$\mathrm{MVN}(\mu, \Sigma) \gets$ \textbf{Training}}\\
	\For {$\textit{sample point i }$}{
		{$F_{i}  \gets$  feature extraction Alg. \ref{FtrAlg}}\\
		{$\mathrm{sc_{MVN}^{i}} \gets$ density values of $F_{i}$ on $\mathrm{MVN}(\mu, \Sigma)$}
	}
	
	\setcounter{AlgoLine}{0}
	\textbf{K-fold Cross Validation:}\\
	{$F_{i}  \gets$  feature extraction Alg. \ref{FtrAlg}}\\
    \For {$\textit{each dataset i }$}{
		\For {$\textit{The remaining datasets j(j!=i)}$}{
		{$threshold  \gets$  EER analysis at \textbf{Training}}\\
		$\mathrm{MVN}(\mu, \Sigma) \gets$  EER analysis at \textbf{Training}}
		}
		    {$\mathrm{sc_{MVN}^{i}} \gets$ probability of $F_{i}$ matching to MVN($\mu$, $\Sigma$)} \\
			{$\mathrm{FPR,FNR} \gets$ apply $threshold$ to $\mathrm{sc_{MVN}^{i}}$
	}
	
	\setcounter{AlgoLine}{0}
	\textbf{Real-time Detection: decision making procedure}:\\
	{$\mathscr{F} \gets$ extract average features for $n$ sample points, Alg. \ref{FtrAlg}}\\
	{$\mathrm{MVN}(\mu, \Sigma) \gets$  \textbf{Training}}\\
	{$threshold  \gets$  \textbf{Training}}\\
	{$\mathrm{sc_{MVN}} \gets $ density values of $\mathscr{F}$ if generated from $\mathrm{MVN}(\mu, \Sigma)$}\\
	\If {$\mathrm{sc_{MVN}} > \textit{threshold}$}{
		\textbf{Authentic } sample point}
	\Else {
		\textbf{Malicious} sample point
	}
	
\end{algorithm}  % mathscr

\textbf{K-fold Cross Validation:} Using the training-testing procedure explained above, we define an extended threshold which identifies all the genuine GPS signals from the spoofed ones for all but a specific dataset. Next, we validate this threshold by calculating the MVN scores for that dataset, and perform the detection process using this threshold. Then, we look into the FPR and FNR to evaluate how well this threshold functions on this dataset which was not included during the training phase. This procedure is repeated for all K-1 datasets, and is called K-fold cross validation \cite{bishop2006pattern}. 

\textbf{Real-time Detection:} The offline defined thresholds in the previous steps are used to trigger the spoofing detector. The receiver generates the features for each sample point and the MVN score of these features based on the distribution of genuine satellite signals. If this score falls below the threshold, the sample will be dropped, otherwise it will be passed over to the next GPS receiver blocks. 

Fig.~\ref{fig::agc} gives an overview of defense for Spotr, and Algorithm \ref{alg::detection} explains the details of steps mentioned above.

\subsection{Time Complexity Analysis}\label{sec::complexity}
We use a very simple feature extraction algorithm with all of the operations performed in time domain which eliminates computational complexity. The feature extraction given at Alg.~\ref{FtrAlg} imposes time complexity of $O\left(1\right)$. Training an MVN model using the training sample points requires covariance matrix calculations and matrix inversion operations with time complexity of $O\left(n^3\right)$. However all this can be done offline. This cuts down the complexity of our algorithm to fitting a single sample point into an already trained MVN model after feature extraction to $O\left(1\right)$.

\section{Experimental Setup}\label{sec::data}
In this section we briefly discuss the organization of GPS navigational messages, followed by the details of our experimental setup and method of data collection. Our datasets, summarized in Tables \ref{table::datasets1} and \ref{table::datasets2}, encompass variations over many months and environmental conditions. The tables contain specifications of the genuine and spoofed datasets, respectively.
\subsection{GPS Data Segmentation}
A GPS transmission is broken into 5 sub-frames that update at varying intervals with an average duration of \unit[6]{s} each. The first sub-frame contains information on the health and accuracy of the satellite, GPS timing information, and any clock corrections.  The second and third sub-frames contain ephemeris data (orbital measurements) for the transmitting satellite.  The fourth sub-frame contains abbreviated almanac data for satellites 25 through 32, as well as ionospheric and UTC data, satellite configuration, and any special messages. The fifth and final sub-frame contains abbreviated almanac data for satellites 1 through 24 and the time and week number of the almanac itself.  Note that sub-frames four and five require approximately \unit[12.5]{mins} to complete, which means sub-frames one, two, and three will update many times during this period. For the attacker to spoof a successful PVT solution, continuous spoofing of between \unit[18]{s} and \unit[30]{s} is required. See the GPS governing agency \cite{U.S.A.F.GPS:System} or \cite{2017UnderstandingApplications,Groves2013PrinciplesSystems,2015GNSSCountermeasures} for more details. 

\subsection{Experimental Hardware}
\begin{figure}[b]
		\centering
		\begin{subfigure}[t]{0.152\textwidth}
		\centering
		\includegraphics[width=\textwidth]{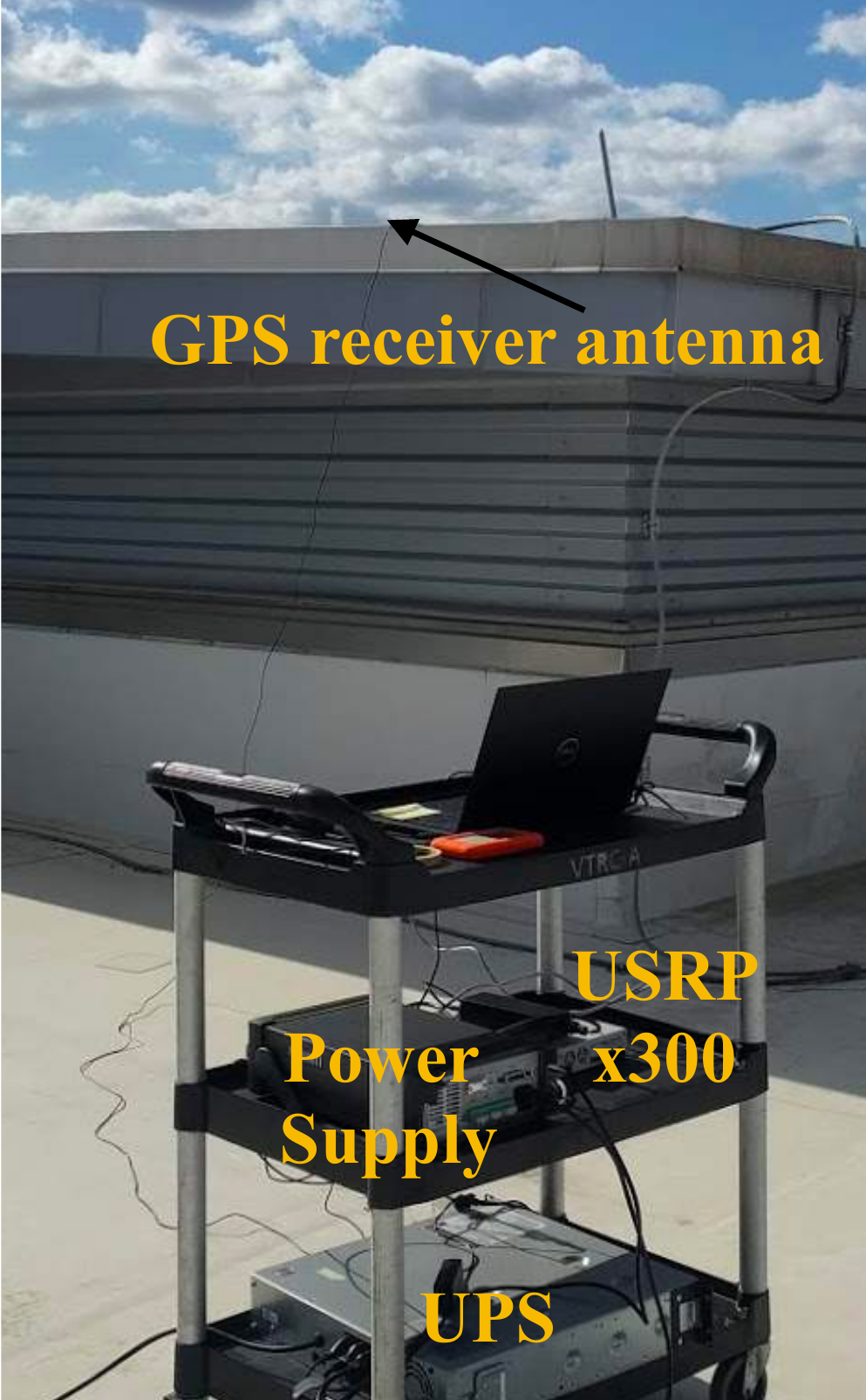}
		\caption{}
		\end{subfigure}
		\begin{subfigure}[t]{0.15\textwidth}
		\centering
		\includegraphics[width=\textwidth]{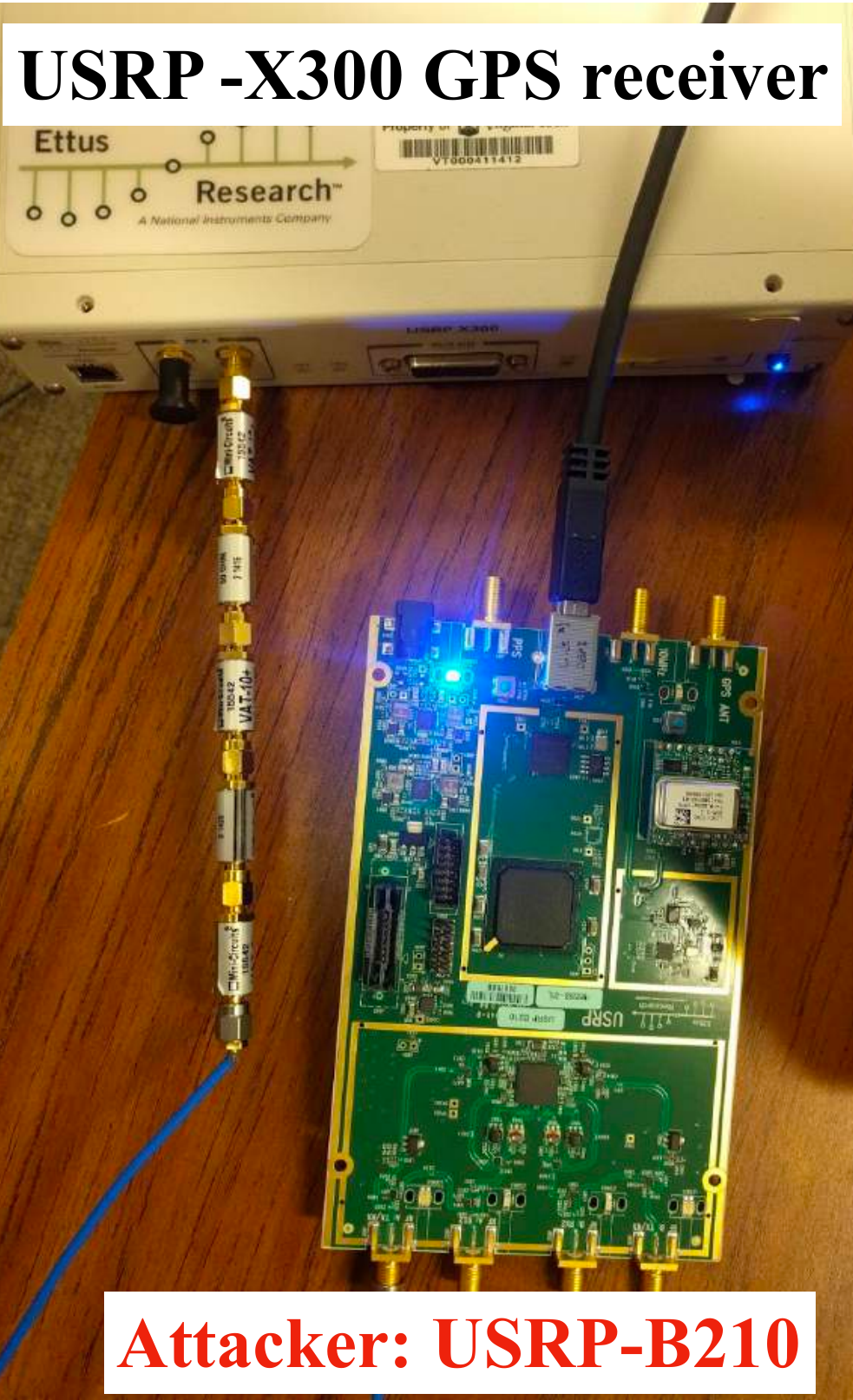}
		\caption{}
		\end{subfigure}
		\caption{(a) Genuine data collection setup (b) Spoofing setup. An Ettus X300 USRP is served as the GPS device fingerprinter, and a B210 USRP for generating the spoofing data.}
		\label{fig::setup}
\end{figure}

We have live data and replay data collections. Our receivers are the B210, the B205mini, and the X300 software-defined radios (SDRs) from Ettus Research.  For live signal collection we use the  X300, which includes a GPS-Disciplined Oscillator (GPSDO) and a UX-160 daughterboard. Two inexpensive (\~\$10) GPS antennae are used, one connected to the RX2 port of the daughterboard and one connected directly to the GPSDO.  For our replay (spoofing) sessions we use the B210 (USB connection) as the transmitter and the X300 (gigabit Ethernet connection) as the receiver.  The two radios are connected (See Fig.\ref{fig::setup}) via SMA-terminated coaxial cabling that begin at the TX/RX port of the B210, pass through a varying attenuator, and end at the RX2 port of the X300. Wired connections like this offer an ideal transmission for the attacker because the channel propagation effects do not influence the signaling of the attacker's radio.     %We accept this criticism, and contend that it was the best course of action given broadcast restrictions over GPS and our desire to prove the concept before proceeding. 

\subsection{Experimental Software}
Our data collection software is GNSS-SDR \cite{Fernandez-PradesGNSS-SDR}, an open source application for GNSS research using SDRs. Signal parameters are set via a configuration file which allows us to specify parameters such as carrier frequency, sampling rate, and data type.  Most importantly, we can arrange for each of the eight available channels to lock only with a specified PRN, ensuring that we can single out each satellite for individual tracking. PRNs are selected based on optimum viewing angle (e.g. as close to directly overhead as possible) and their positions are correlated between an online tracking website (https://in-the-sky.org) and an Android app (GPS Test) \cite{GPStest}.

GPS-SDR-SIM \cite{gpssim}, another open source application, is our primary means of both generating rudimentary sets of spoofed data and for transmitting said data over the wired connection.  This application allows one to present a GPS ephemeris file (via a repository maintained by NASA), specify time, date, location, and duration variables, and then generate authentic GPS binary data ready for transmission. One issue we encountered is that while the software does generate viable GPS NAV messages and we can successfully spoof a location with them, the software is not meant to address PRN authenticity. As such, we modify this application to allow for greater specificity in the data generated resulting in a spoofed signal accurate in time, location, and PRNs received. Our modifications will be shared for use by the security community.

A summary of the genuine and spoofed GPS data (SatGrid) that we collected using our setup is given in Tables \ref{table::datasets1} and \ref{table::datasets2}, respectively. The collection date for the spoofed data corresponds to the date that the genuine data was acquired, not when it was actually replayed. We discuss this in more detail in Sec. \ref{sec::rslt}. %Nomenclature of table same as \cite{humphreys2012texas}.  which we would be glad to share by the community.

\subsection{Second-Party Data: UT-Austin TexBat Repository}\label{sec::TexBat}
Several works in the field of GPS security are tested against de-facto standard of a publicly available repository of GPS signal spoofing traces called Texas Spoofing Battery (TexBat) \cite{humphreys2012texas}. The dataset is provided by the Radionavigation Lab at Univ. of Texas-Austin, including 10 rounds of data collection with a duration of \unit[400]{s} each where the first \unit[100]{s} being spoof-free, sampled at \unit[25]{Msps} with a \unit[16]{bit} resolution for complex values. TexBat includes two rounds of separate spoof-free data, which are categorized based on mobility of the platform as \textit{clean static} and \textit{clean dynamic}. Clean static data collected from a reference antenna located in a building, while the clean dynamic data is recorded from an antenna
mounted atop a vehicle traveling in Austin, TX. The rest of the datasets, are spoofed data with different attacks in time and/or position \cite{humphreys2012texas} which have as low as meter level alignment with the genuine signals. TexBat induces a \unit[600]{m} position offset, and \unit[2]{\micro s} error in time. 

In this work we test Spotr on TexBat data in addition to our dataset (SatGrid). First, because it provides a fair comparison platform for security researchers and helps them go beyond statistical analysis methods for testing their solutions. The detection methods which were limited to verify the null hypothesis (no attack) before, were then able to test alternate hypothesis thanks to the spoofing data of TexBat. Second, the UT spoofer that generates the TexBat data (Table \ref{table::datasets2}), is a strong attacker which is diverse in terms of dimensions of the genuine GPS signals that it targets to attack.

\section{Experimental Results}\label{sec::rslt}
In this section we evaluate Spotr on different available attack scenarios operating at different conditions of time, location, multipath richness, and/or hardware.  For reference, Tables \ref{table::datasets1} and \ref{table::datasets2} summarize all genuine and spoofed data respectively, both our own live recordings and the Univ. of Texas-Austin ``TexBat'' \cite{humphreys2012texas} repository.

We follow two main objectives in fingerprinting the GPS signals using the features that we introduced at Sec. \ref{sec::ftrs}. First, to identify a genuine satellite signal from a spoofed one, referred to as \textit{detection}. Second, to prove the consistency of these features by tracking the signals from a genuine satellite on different conditions, called \textit{tracking}. We evaluate the detection and tracking capability of Spotr on different types of attacks on GPS signals listed at Table \ref{table::datasets2}. First we evaluate a spoofing attacker over different environmental conditions in Sec.\ \ref{eval::inept}. Next, Spotr is evaluated against the stronger replay attacker with more insights on the success rate of the attacker in inducing falsified PVT solutions in Sec.\ \ref{eval::adept}. We cannot compare the error rates with state-of-the-art, because they report results using higher level metrics, such as the maximum location offset \cite{Ranganathan2016SPREE:Receiver, broumandan2015spoofing}. See Sec. \ref{Sec::relatedwork} for details on SPREE which limits the range of spoofing attacks on position to the radius of \unit{1000} {m}, whereas we report maximum continuous spoofing time of \unit[47.3]{s} in Sec.\ref{eval::adept}. Also, our low error rates are indicative of the fact that circuitry of a satellite is different from an SDR (D1) or non-SDR (D2,D3) spoofer.

For the scenarios in Table \ref{table::xval} the training and testing column include the general dataset information, followed by the genuine datasets from device \textit{m} (Dm) labeled by Dm:Gx and spoofed datasets collected from Dm labeled by Dm:Sx, with x being the dataset index in the Tables \ref{table::datasets1} and \ref{table::datasets2}. For the cases that only one dataset is reported (e.g. TexBat:S5) the first \unit[100]{s} is spoof-free.

Note that the low values on the Y-axis of all the figures are due to the high dimensionality of the data (six). The MVN density values close to the maximum achievable MVN score are interpreted as high scores. For example, the maximum achievable MVN score of Fig.~\ref{fig::time1} is 6e-27 which is calculated by using a given $\Sigma$ of the training data, $k=6$ and $x=\mu$ with Eq.\ \ref{eq::mvn}.

\subsection{Evaluation and Discussion}\label{eval::inept}
We evaluate Spotr against spoofing attacker over different times, locations, multipath conditions, and the fingerprinter device being used in the data collection setup, in the following. 
\subsubsection{Detection and Tracking over Time}\label{time}
After generating features based on Sec. \ref{sec::ftrs}, we randomly select 40,000 samples from SatGrid:G1 and SatGrid:S1 (Spoofing-Attacker) on Sep 24, 2018 in \LocOne{} and fit an MVN model using Eq. \ref{eq::mvn} to train a  $\mu$ and $\Sigma$. Next, we use the remaining data on the same day to do EER analysis on the MVN scores, and train an identification/detection threshold. Note that, as many sample points as needed (shown by \textit{n}) to achieve an EER of zero (the ideal value for EER) are used in all of the cases, which is reported in the X-axis label of the graphs if required. 

 Next, we use the trained MVN model and the threshold to test the genuine and spoofed data collected on different days using the same spoofer at \LocOne{} in consecutive days right after the training, also after one year on Sep 10, 2019 at \LocTwo{} (SatGrid:G23). Fig.~\ref{fig::time1} shows the MVN scores associated with this analysis. We observe that without any averaging on the sample points, all of the genuine data have MVN scores above the trained threshold while the MVN scores of spoofed data are mostly zero (which is the reason they are not printed in the graph with logarithmic scale on Y-axis). This allows us to conclude that we are able to track a genuine signal over the period of one year which validates the stability of our features with time.  Scenario 2 of the Table \ref{table::xval} illustrates the error rates (FNR and FPR) as well as the required number of averaged sample points (n) to achieve the reported error rates for all the data of \LocOne{} and  \LocTwo{} collected on Sep 2018 and 2019. Table \ref{table::kfoldSatGrid} shows the results of a more general training-testing procedure on SatGrid data collected on Sep 2018 using K-fold cross validation.
 
 Similar analysis is performed at Scenario 7 of Table \ref{table::xval} on a selection of TexBat data to validate the stability of our features over time. Unlike Scenario 2, both the data used for training and testing in this case are collected in rich-multipath environment. Fig.~\ref{fig::time2} shows the MVN scores of the test data, TexBat:S6, where the MVN model and the threshold are trained on TexBat:S5. Presence of strong multipath is the reason for the need to average multiple sample points (35000 in this case), before making a decision on their authenticity. 

\subsubsection{Detection and Tracking at Different Locations}\label{location}
Following the same training process in Sec.~\ref{time}, we test the detection and tracking of Spotr on the data collected from different locations using spoofing attacker given as Scenario 1 in Table \ref{table::xval}. Fig.~\ref{fig::location} shows MVN scores of the data collected in Missouri, while the original MVN model and threshold are trained based off of \LocOne{}. Using detection threshold of 1.77e-92 for all of the satellites that appeared on different days of data collection listed at Table \ref{table::datasets1}, leads to EERs of zero, which allows us to conclude that we can distinguish the genuine signals from spoofed ones regardless of the location.

\subsubsection{Detection and Tracking in Presence of Multipath}\label{path}
The most common challenge in spoofing detection of GPS systems is the multipath effect which makes it difficult to distinguish a genuine multipath component of GPS signal from the malicious ones. This problem is elaborated by Wesson et. al \cite{wesson2011evaluation}, and stated as a limitation at \cite{ranganathan2016spree,montgomery2011receiver}, however a solution has not been proposed yet. To investigate the performance of Spotr in presence of multipath we fit an MVN model to the multipath-free training data, and test it on rich-multipath data and vice versa for both SatGrid and TexBat (Scenarios 3, 7, 8 and 9 of Table \ref{table::xval}). This round of analysis can help us understand the extent to which genuine data with multipath effect can be confused with spoofing data.% if this propagation effect is not taken into consideration at training stage. 
 
The initial set of TexBat data is collected on 2012 using their first fingerprinting hardware (D1) on a static platform with line of sight among which TexBat:S1, S2, S3, and D4 are the multipath-free ones. Another set from the TexBat repository is collected in a rich-multipath environment with the same fingerprinter mounted on a vehicle and driven through a populated area of Texas for a 3 mile. TexBat:S5, and S6 are collected from this ``dynamic'' platform. Fig.~\ref{fig::mp_texbat_dev1} illustrates the MVN scores of genuine and spoofed data where model is trained on this rich-multipath data and tested on the earlier mentioned multipath-free datasets from TexBat D1. Scenario 9 of Table \ref{table::xval} reports the error rates of this analysis with a FNR of $5.23\%$ which means there is a spoofing activity which remains undetected for this case after using 1000 sample points for a final decision making on their authenticity. As mentioned earlier at Sec. \ref{sec::digreceiver} the VOLK-Library \cite{Fernandez-PradesGNSS-SDR} from gnss-sdr that is responsible for running the dll-pll loops at the GPS receiver, runs at a varying speed in a feedback system. This is why we cannot translate the number of undetected malicious sample points to the notion of time and report the maximum continuous time that a spoofing activity will remain undetected by Spotr without access to accurate timestamps. We elaborate on this timing problem in Sec.~\ref{eval::adept} by analyzing the worst case scenario (strongest attacker: matched-powered replay attacker on SatGrid) with timestamps.   

To further evaluate the robustness of Spotr for all possible cases, we perform training and testing on the rich-multipath data from TexBat in Scenario 7 of Table \ref{table::xval}, where using multiple sample points helps us to overcome the influence of multipath on the detection/tracking process shown in Fig.~\ref{fig::time2}. We also evaluate Spotr on SatGrid at Scenarios 3 and 8, where training is done on a multipath-free (Sep 10, 2019) and rich-multipath data (Aug 23, 2019), respectively. We are able to perform highly accurate detection/tracking with only one sample point as demonstrated in Fig.~\ref{fig::mp} and Fig.~\ref{fig::no_mp}.

\subsubsection{Detection and Tracking against Different Attacks}\label{fingerprinter}
Table \ref{table::kfold_texbat_dev1} shows the results of a more general training-testing procedure on TexBat data collected using D1 2012 using K-fold cross validation where the spoofing types vary from single time or position, to simultaneous time and position detailed in Table \ref{table::datasets2}. This analysis is performed on a mix of multipath-free data (Fig.~\ref{fig::no_mp_texbat_dev1} and Fig.~\ref{fig::no_mp_texbat_dev2}) and data collected in rich multipath (Fig.~\ref{fig::mp_texbat_dev1}) at the same time. The error rates in the table indicates high FPR and FNR values in some occasions, caused by using the multipath affected data at training phase. This shows that genuine data with multipath effect can be confused with spoofing data if this propagation effect is not taken into consideration at training stage. To compensate for the reduced accuracy in the detection/tracking process, we increase the number of observation sample points and report new rates in Table \ref{table::kfold_texbat_dev1_n}. 

\subsubsection{Detection and Tracking using Different Hardware Platforms as Fingerprinter}\label{fingerprinter}
\begin{figure*}[thb!]
	\centering
	\begin{subfigure}[t]{.33\textwidth}
		\centering
		\includegraphics[height=1.75in,width=0.9\columnwidth]{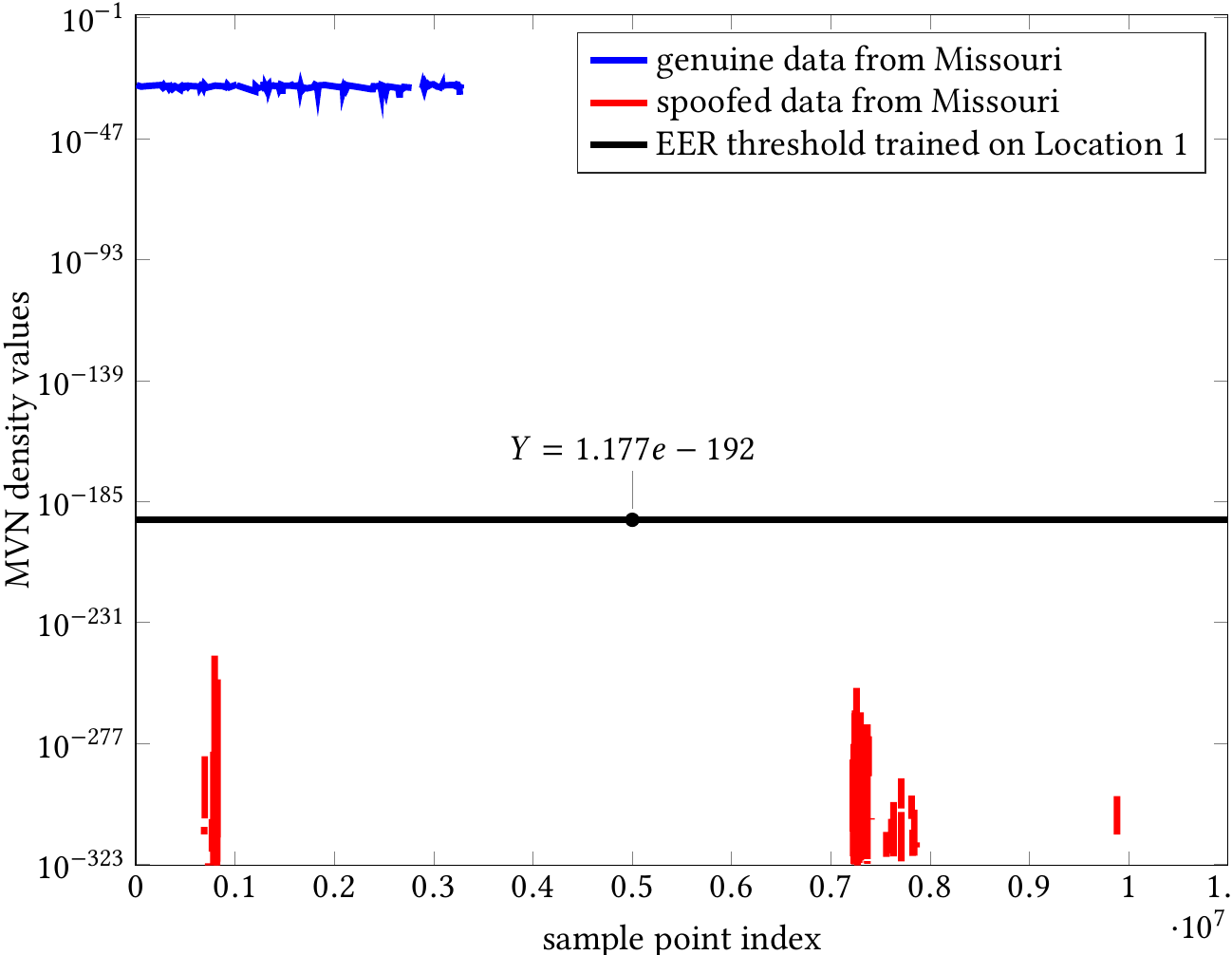}
		\caption{}
		\label{fig::location}
	\end{subfigure}
	\begin{subfigure}[t]{.33\textwidth}
		\centering
	\includegraphics[height=1.75in,width=0.9\columnwidth]{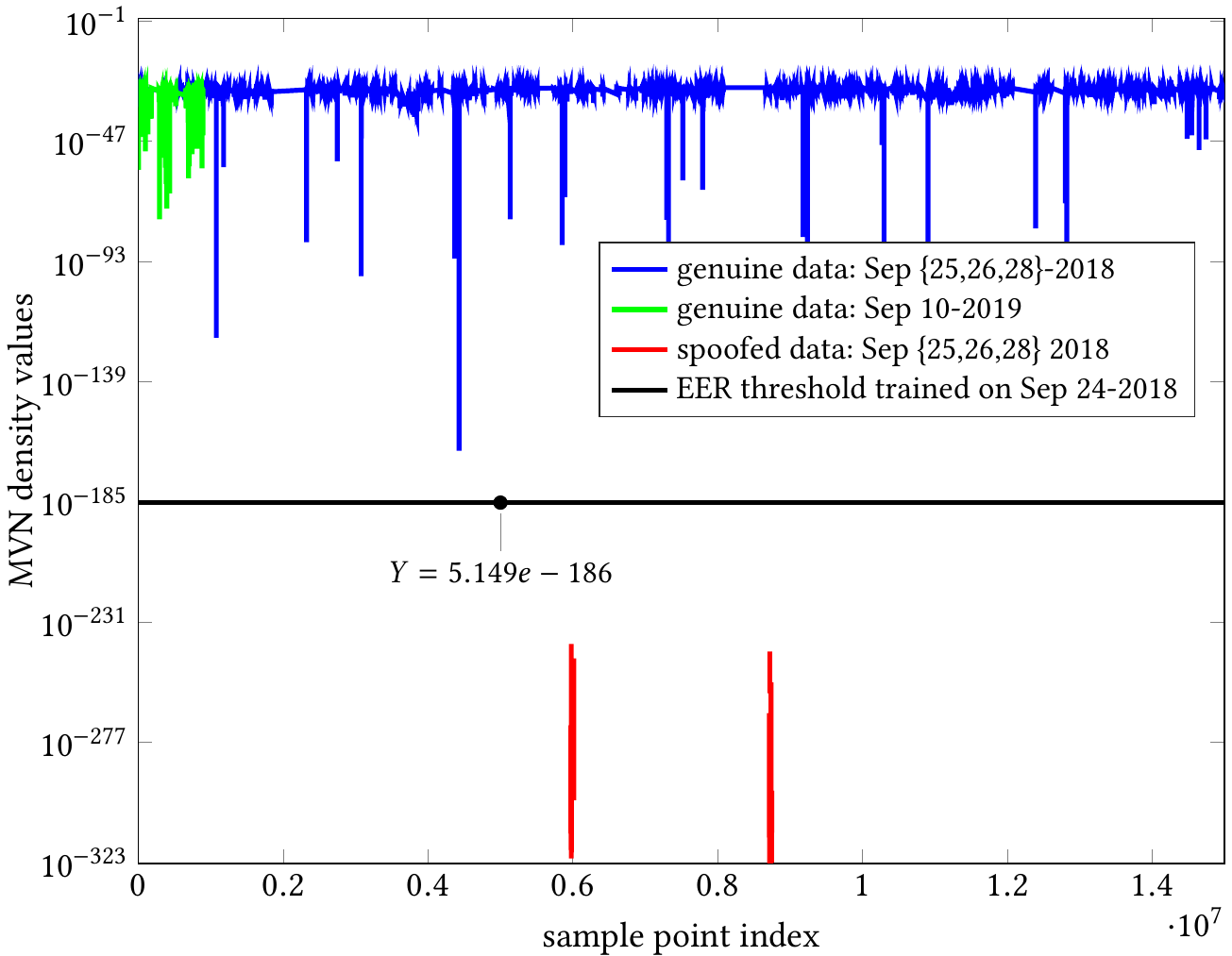}
		\caption{}
		\label{fig::time1}
	\end{subfigure}
	\begin{subfigure}[t]{.33\textwidth}
		\centering
		\includegraphics[height=1.75in,width=0.9\columnwidth]{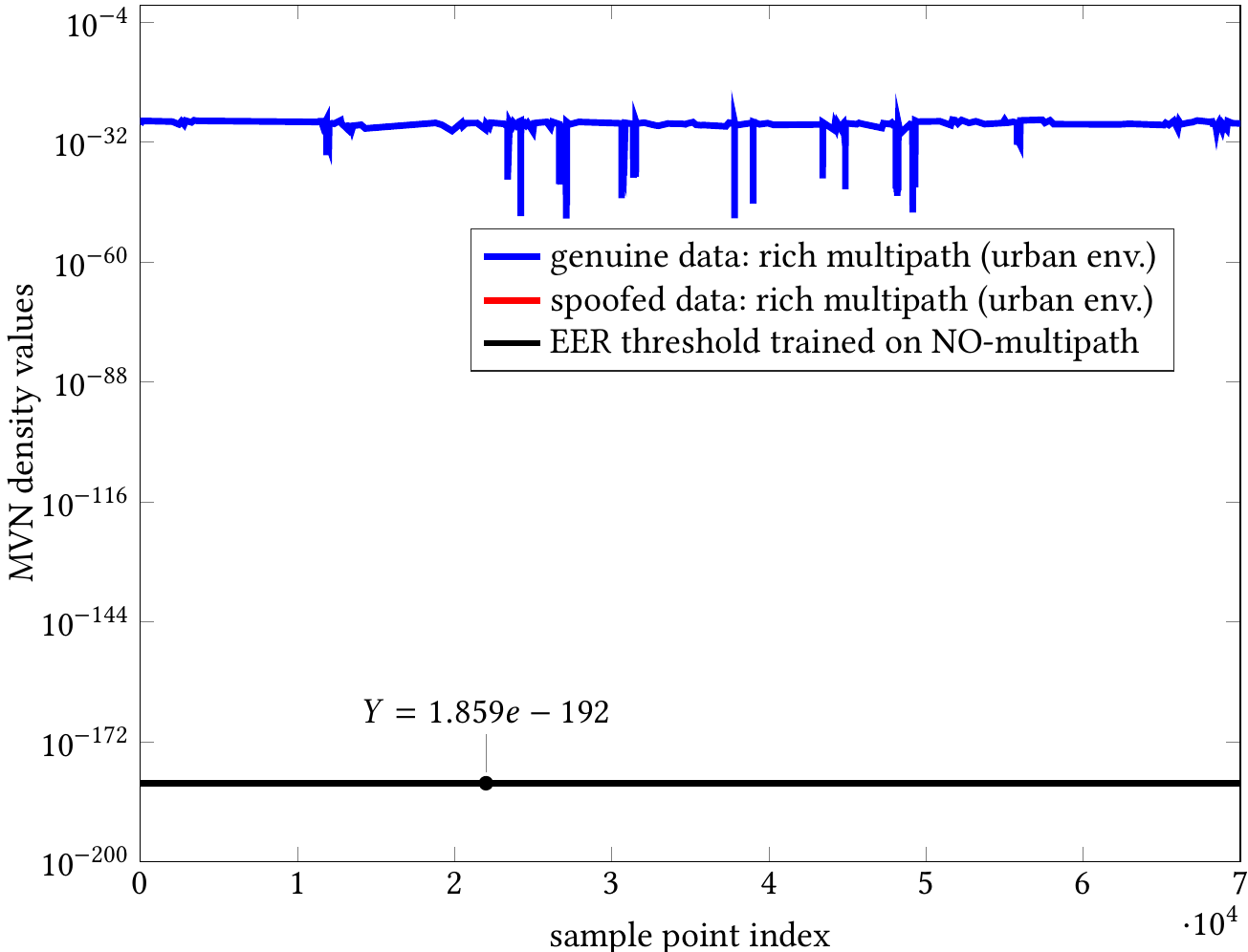}
		\caption{}
		\label{fig::mp}
	\end{subfigure}
	\begin{subfigure}[t]{.33\textwidth}
		\centering
		\includegraphics[height=1.75in,width=0.9\columnwidth]{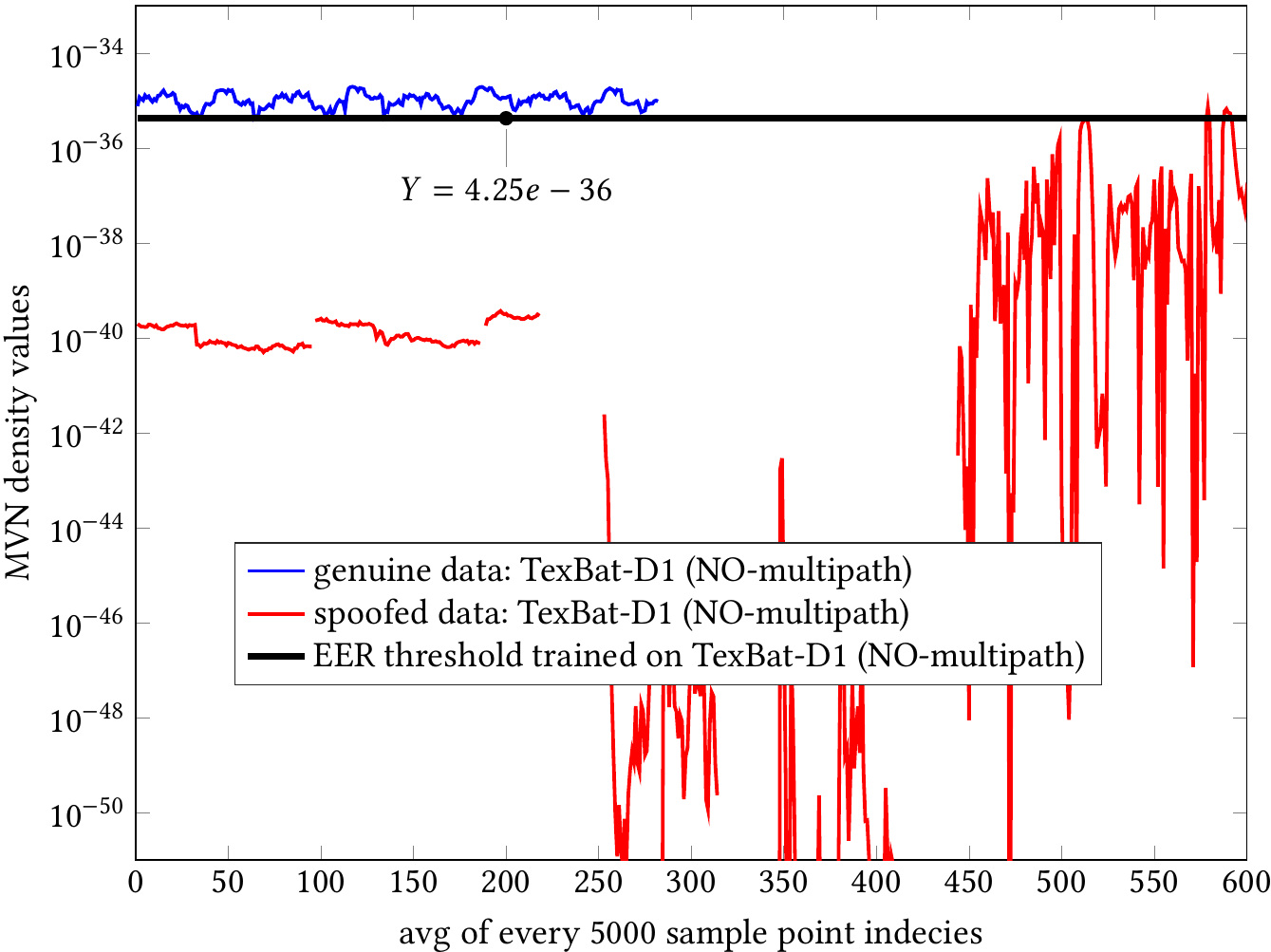}
		\caption{}
		\label{fig::no_mp_texbat_dev1}
	\end{subfigure}
	\begin{subfigure}[t]{.33\textwidth}
		\centering
		\includegraphics[height=1.75in,width=0.9\columnwidth]{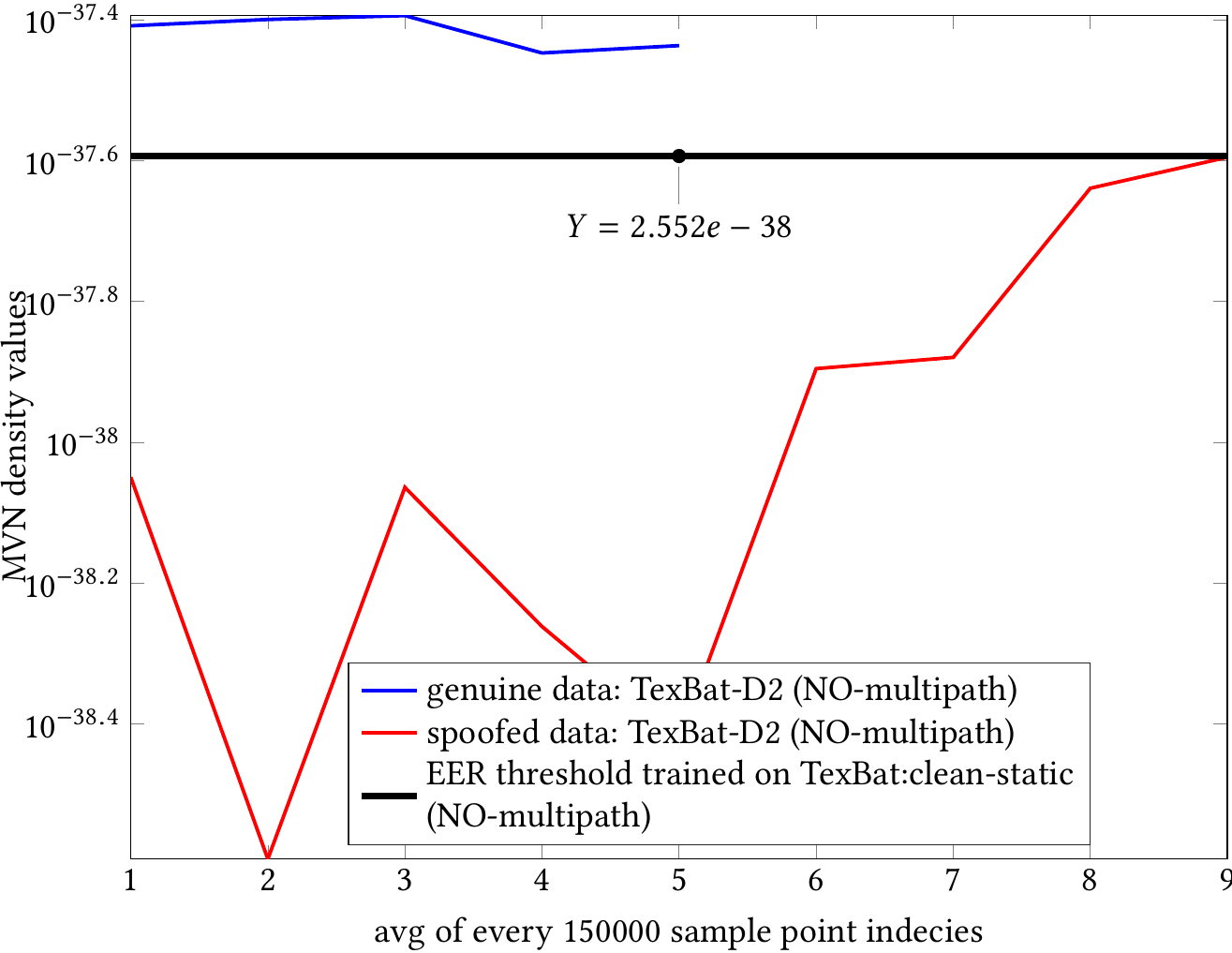}
		\caption{}
		\label{fig::no_mp_texbat_dev2}
	\end{subfigure}
	\begin{subfigure}[t]{.33\textwidth}
		\centering
		\includegraphics[height=1.75in,width=0.9\columnwidth]{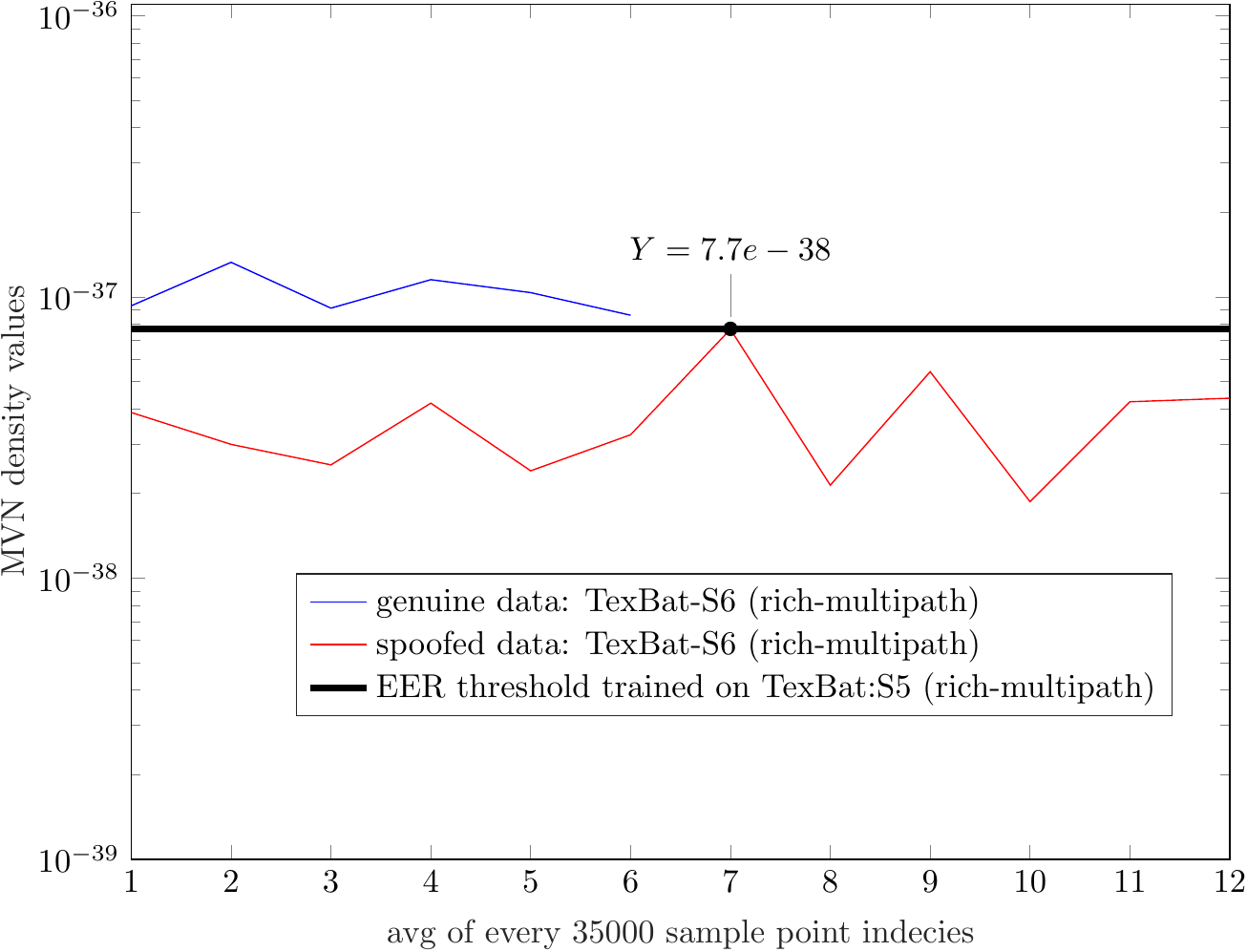}
		\caption{}
		\label{fig::time2}
	\end{subfigure}
	\begin{subfigure}[t]{.33\textwidth}
		\centering
		\includegraphics[height=1.75in,width=0.9\columnwidth]{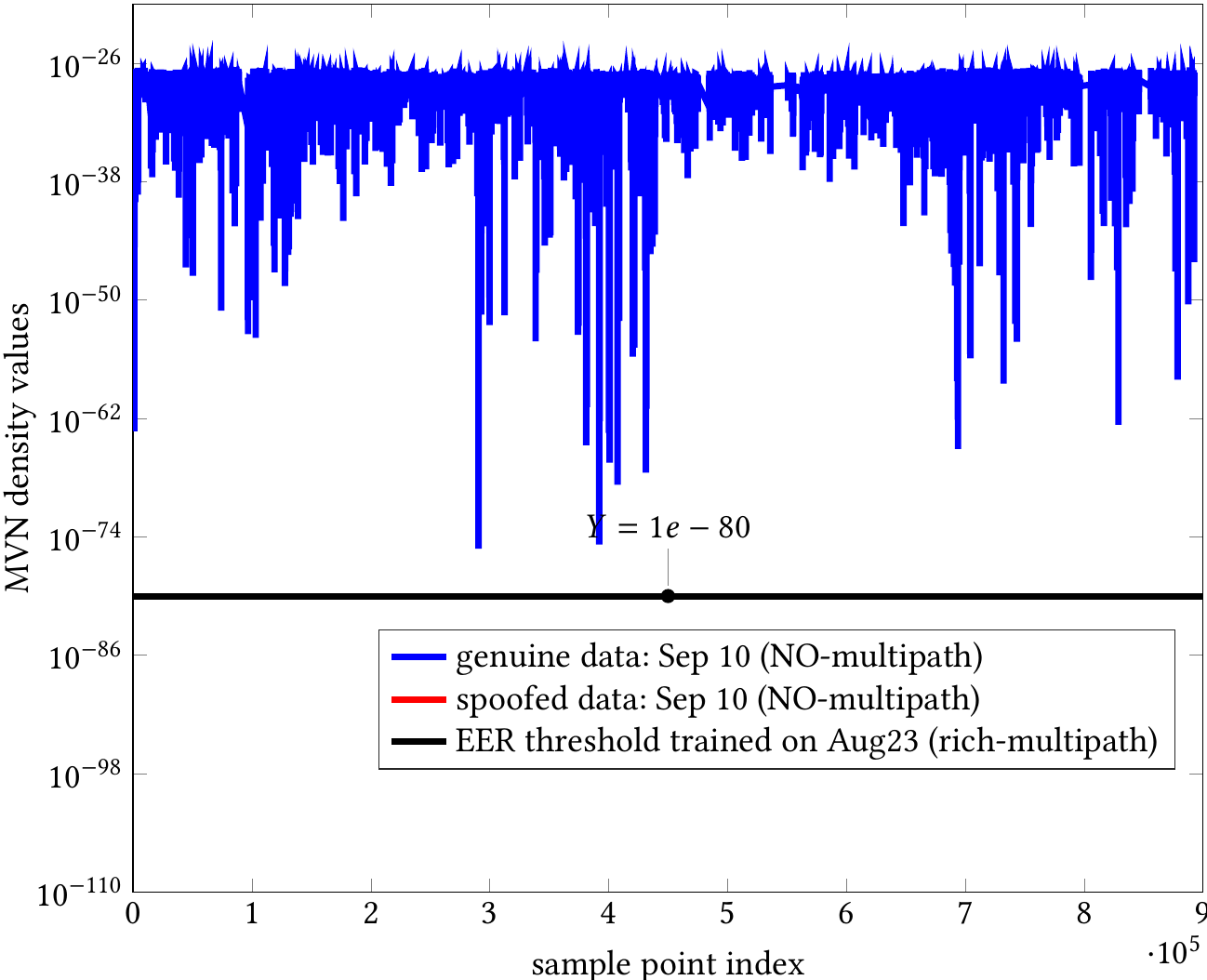}
		\caption{}
		\label{fig::no_mp}
	\end{subfigure}
	\begin{subfigure}[t]{.33\textwidth}
		\centering
		\includegraphics[height=1.75in,width=0.9\columnwidth]{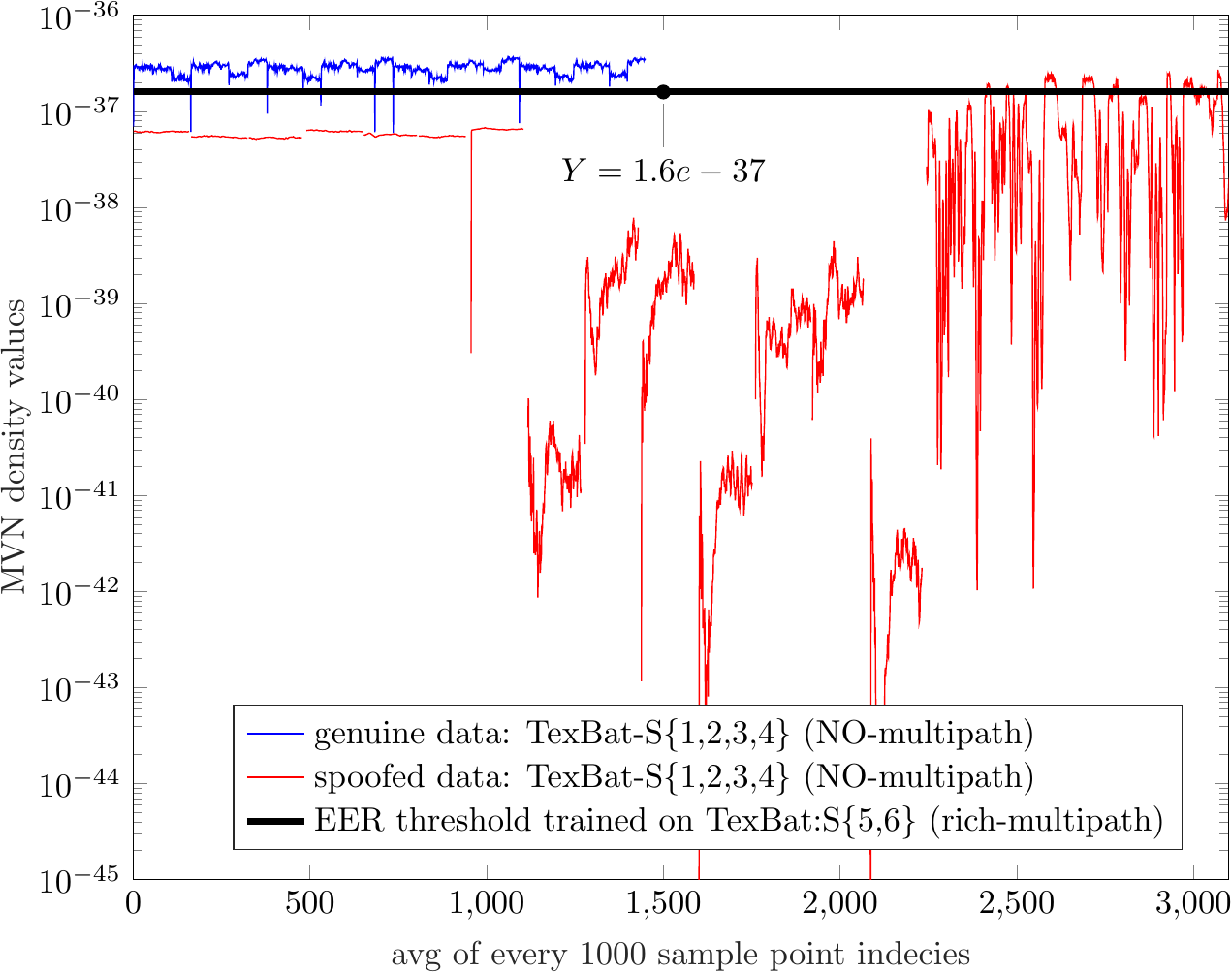}
		\caption{}
		\label{fig::mp_texbat_dev1}
	\end{subfigure}
		\begin{subfigure}[t]{.33\textwidth}
		\centering
		\includegraphics[height=1.75in,width=0.9\columnwidth]{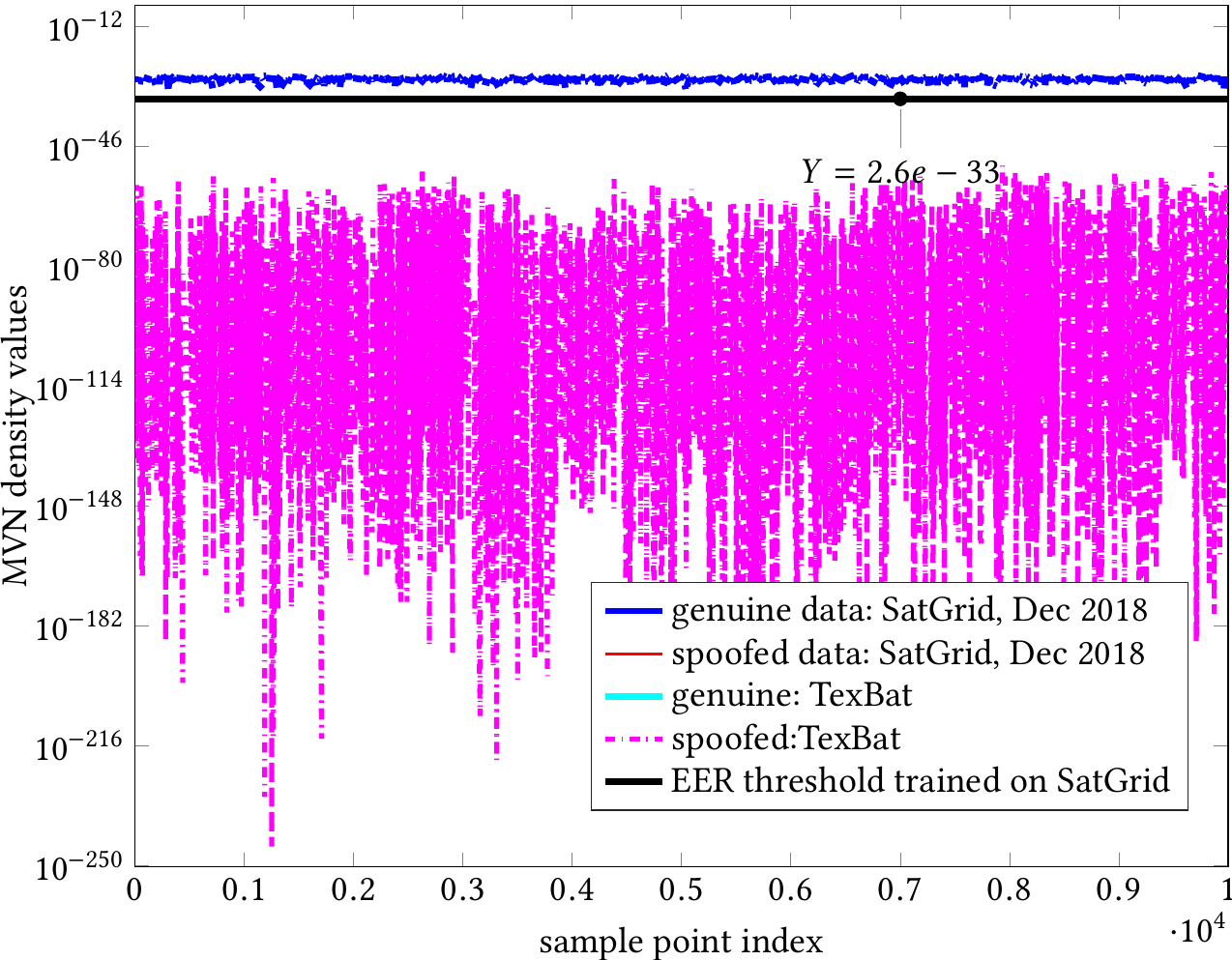} % line 6
		\caption{}
		\label{fig::fingerprinter}
	\end{subfigure}
	\caption{The results of using Spotr for detection/tracking of spoofed/genuine signals (a) cross locations (b)-(f) cross time (c)-(g) in the presence or absence of multipath (d)-(e)-(h) for spoofing attacks of TexBat using different hardware platforms (i). At Figures (a)-(b)-(c) and (g), most of the MVN scores for spoofed signals (depicted by red line) are exact zero values, hence not printed in the graph with logarithmic scale on Y-axis. Logarithmic scale is used for better visualization. }\label{fig::rslts}
\end{figure*}

The main idea behind our proposed spoofing detection is hardware fingerprinting of the transmitters. This allows us to hypothesize that the features that we exploit in our algorithm should change with the signal acquisition platform. In this section, we validate this by training a model on the data from SatGrid and testing it on the data from TexBat. Fig.~\ref{fig::fingerprinter} shows that we are not able to track the genuine data from TexBat fingerprinting hardware when training a model with SatGrid (the MVN density values of the genuine data from TexBat all hold value of zero, that is why they are not printed in logarithmic scale of Y-axis). This is a general limitation of fingerprinting approaches (even if the platforms are identical \cite{foruhandeh2019simple}) and can be overcome by training each fingerprinter on genuine satellite data before deployment.

\begin{table*}[t!]
\caption{This table demonstrates the efficacy of Spotr for detection/tracking of spoofed/genuine signals (listed in Tables \ref{table::datasets1} and \ref{table::datasets2}) across locations, times, in the presence or absence of multipath (MP), and for different attacks using different hardware platforms (TexBat:D1 and D2 or SatGrid:D3). The table reports thresholds for obtaining equal error rates (EER) of zero at the training phase, as well as false positives (FPR) and false negatives (FNR) with the number of required sample point observations (n) for attaining the reported FPR and FNR at the testing phase.  \textit{key}: \textit{The training and testing column include, the general dataset information, followed by the genuine datasets from device m (Dm) labeled by Dm:Gx and spoofed datasets collected from Dm labeled by Dm:Sx, with x being the dataset index in the Tables \ref{table::datasets1} and \ref{table::datasets2}. For the cases that only one dataset is reported (TexBat, D1 2012) the first \unit[100]{s} of the data is spoof-free. For example, D3:G1,G2,G3,G4 shows the genuine data collected from SatGrid, indexed by G1,G2,G3 and G4 in Table \ref{table::datasets1}.}}\label{table::xval}
\small
\begin{tabular}{l|c|c|c|c|c|c|c}
Scenario & Train & Threshold  & Test & FPR & FNR & n & Plot\\
\hline 
\hline
\rowcolor{blue!0}
 \rownumber  $.$  \, Location & \begin{tabular}[c]{@{}c@{}} SatGrid (\LocOne{},$\sim$MP) \\D3:G1,G2,G3,G4 \\ D3:S1,S2,S3,S4 \end{tabular}  &1.177e-192  & \begin{tabular}[c]{@{}c@{}}SatGrid (Missouri,$\sim$MP) \\ D3:G5,G8 \\ D3:S5,S6 \end{tabular} & 0\% & 0\%  & 1 & Fig. \ref{fig::location}
\\ % line 1

  \rownumber  $.$  \, Date & \begin{tabular}[c]{@{}c@{}} SatGrid (\LocOne{},$\sim$MP) \\ D3:G1 \\ D3:S1 \end{tabular}  & 5.15e-186  &  \begin{tabular}[c]{@{}c@{}} SatGrid (\LocOne{},$\sim$MP) \\ D3:G2,G3,G4 \\ D3:S2,S3,S4 \end{tabular} & 0\% & 0\%  & 1 & Fig. \ref{fig::time1}
\\ % line 2
 
\rowcolor{blue!0}

 \rownumber  $.$  \, MP & \begin{tabular}[c]{@{}c@{}} SatGrid (\LocTwo,$\sim$MP) \\ D3:G23 \\ D3:S8 \end{tabular}  & 1.86e-182   & \begin{tabular}[c]{@{}c@{}} SatGrid (\LocTwo,MP) \\ D3:G22 \\ D3:S7\end{tabular} & 0\% & 0\% & 1 & Fig. \ref{fig::mp} \\ % line 3 
 
  \rownumber  $.$  \, Attack (D1) &  \begin{tabular}[c]{@{}c@{}} TexBat (Texas,$\sim$MP) \\ D1:S1,S2,S3,S4 \end{tabular} & 4.25e-36 &  \begin{tabular}[c]{@{}c@{}} TexBat (Texas,$\sim$MP) \\ D1:S1,S2,S3,S4 \end{tabular}  & 0\%  & 0.96\% & 500 & Fig. \ref{fig::no_mp_texbat_dev1} \\ % line 4
 
\rowcolor{blue!0}

 \rownumber  $.$  \, Attack (D2) &  \begin{tabular}[c]{@{}c@{}} TexBat (Texas,$\sim$MP)\\ D2:Clean1 \\ D2:S7,S8 \end{tabular} & 2.56e-38  &  \begin{tabular}[c]{@{}c@{}} TexBat (Texas,$\sim$MP) \\ D2:Clean1 \\ D2:S7,S8 \end{tabular}  &  0\% & 0\% & 15000 & Fig. \ref{fig::no_mp_texbat_dev2} \\ % line 4-2

 \rownumber  $.$  \, Hardware &  \begin{tabular}[c]{@{}c@{}} SatGrid (Missouri) \\ D3:G5,G8 \\ D3:S5,S6 \end{tabular} & 2.6e-33 &  \begin{tabular}[c]{@{}c@{}} TexBat (Texas,$\sim$MP) \\ D1:S1,S2,S3,S4 \end{tabular}  &  100\% & 0\% & 1 & Fig. \ref{fig::fingerprinter} \\ % line 6

 \hdashline[1.5pt/5pt]
  \rowcolor{blue!0}
  \rownumber  $.$  \, Date & \begin{tabular}[c]{@{}c@{}} TexBat (Texas,MP) \\ D1:S5 \end{tabular}  & 7.7e-38  & \begin{tabular}[c]{@{}c@{}} TexBat (Texas) \\ D1:S6 \end{tabular}  & 0\%  & 0\% & 35000 & Fig. \ref{fig::time2} \\  % line 7

  \rownumber  $.$  \, $\sim$MP & \begin{tabular}[c]{@{}c@{}} SatGrid (\LocTwo,MP) \\ D3:G22 \\ D3:S7 \end{tabular}   & 1e-80 & \begin{tabular}[c]{@{}c@{}} SatGrid (\LocTwo,$\sim$MP) \\ D3:G23 \\ D3:S8 \end{tabular} & 0\%  & 0\% & 1 & Fig. \ref{fig::no_mp} \\ %line 8
 
  \rowcolor{blue!0}
 \rownumber  $.$  \,  Attack (D1) &  \begin{tabular}[c]{@{}c@{}} TexBat (Texas,MP) \\ D1:S5,S6 \end{tabular}  & 1.6e-37  &  \begin{tabular}[c]{@{}c@{}} TexBat (Texas,$\sim$MP) \\ D1:S1,S2,S3,S4 \end{tabular} & 0\%  & 5.23\% & 1000 & Fig. \ref{fig::mp_texbat_dev1} \\  % line 9
\end{tabular}

\end{table*}

\begin{table*}[t]
\small
\centering
\caption{ (a) Evaluation of detection process for the genuine (G1-G4) and spoofed (S1-S4) data generated by SatGrid (D3) using FPR and FNR values. (b) 6-fold cross validation on TexBat spoofing scenarios (S1-S6) including both multipath/non-multipath data for the 2012 fingerprinter (D1) based on one sample point of observation. (c) 6-fold cross validation on TexBat spoofing scenarios (S2-S6) based on 40,000 sample points of observations.}
% naiive attacker results
\begin{subtable}{0.34\linewidth}
\caption{}\label{table::kfoldSatGrid}
\tiny

     \begin{tabular}{c|r|r|r}
        \hline
           \thead{Validation \\ Dataset}    & FPR & FNR  & EER threshold \\
     %   \hline
        \hline\hline
        \textbf{$\mathrm{SatGrid:S1 \& G1}$}             & $0\,\%$       & $0\,\%$   & \num{0.6829e-31} \\
        \hline
        \textbf{$\mathrm{SatGrid:S2 \& G2}$}         & $0\,\%$       & $0\,\%$   &\num{0.5126e-31} \\
         \hline
        \textbf{$\mathrm{SatGrid:S3 \& G3}$}         & $0\,\%$       & $0\,\%$   &\num{0.0138e-31} \\
         \hline
        \textbf{$\mathrm{SatGrid:S4 \& G4}$}         & $0\,\%$       & $0\,\%$   &\num{0.8071e-31} \\
        \hline
    \end{tabular}%

\end{subtable}
% TexBat spoofers on clean_static results 
\begin{subtable}{0.34\linewidth}
\caption{}\label{table::kfold_texbat_dev1}
\tiny

\begin{tabular}{c|c|c|c}
\hline
\thead{Validation \\ Dataset}         & FPR & FNR & EER Threshold  \\ 
\hline\hline
\textbf{$\mathrm{TexBat:S1}$} & 0\%          &  0\%          & \num{3.6370e-36}   \\ \hline
\textbf{$\mathrm{TexBat:S2}$} & 0\%          &  0\%          & \num{2.4183e-36}   \\ \hline
\textbf{$\mathrm{TexBat:S3}$} & 0.38\%          & 3.35\%         & \num{9.4367e-36} \\ \hline
\textbf{$\mathrm{TexBat:S4}$} & 0.18\%          & 46.8\%          & \num{9.8347e-37}   \\ \hline
\textbf{$\mathrm{TexBat:S5}$} & 26.29\%          & 0\%          & \num{2.9695e-35}     \\ \hline
\textbf{$\mathrm{TexBat:S6}$} & 24.48\%          & 0\%          & \num{2.5882e-35}    \\ \hline
\end{tabular}

\end{subtable}
\begin{subtable}{0.3\linewidth}
\caption{}\label{table::kfold_texbat_dev1_n}
\tiny
\begin{tabular}{c|c|c|c}
\hline
\thead{Validation \\ Dataset}         & FPR & FNR & EER Threshold  \\ 
\hline\hline
\textbf{$\mathrm{TexBat:S1}$} & 0\%          & 0\%      &  \num{3.544e-36}   \\ \hline
\textbf{$\mathrm{TexBat:S2}$} & 0\%          & 0\%      &  \num{2.4183e-36}   \\ \hline
\textbf{$\mathrm{TexBat:S3}$} & 0\%          & 0\%          &  \num{2.6983e-35}\\ \hline
\textbf{$\mathrm{TexBat:S4}$} & 0\%          & 12.5\%      &  \num{1.8412e-35}  \\ \hline
\textbf{$\mathrm{TexBat:S5}$} & 0\%          & 0\%          &  \num{5.51976e-35}  \\ \hline
\textbf{$\mathrm{TexBat:S6}$} & 0\%          & 0\%          &  \num{3e-35}   \\ \hline
\end{tabular}

\end{subtable}
\end{table*}

\subsection{Security Analysis: Feature Replay}\label{eval::adept}
In this section, we look into the capabilities of our strongest attacker, where we give the attacker exact same samples as that of a genuine satellite. Table \ref{table::datasets1} includes two rounds of SatGrid GPS data collected on Nov 8, 2019, which unlike all other listed datasets includes high fidelity timestamps. Data from a matched-power replay attack associated with this data is also given at Table  \ref{table::datasets2}. This is the strongest attacker amongst the ones listed in Table \ref{table::datasets2} for several reasons. First, the attacker is capable of estimating and matching its power to the power level of the genuine GPS signal at the target receiver in a real-time manner. This represents the best case (unrealistic) scenario for the attacker as it hides the spoofing activities from not only in-band power based spoofing detectors but also  precludes anomalies at the complex correlator output tabs of the receiver caused by the struggle between the genuine and the spoofed signals aiming to take control of the tracking block \cite{humphreys2016texbat}. Second, because of our attack model and data collection setup, some inevitable amount of delay is inherent to the attacker, which we have removed. In literature of device fingerprinting \cite{danev2010attacks}, this is the strongest attacker. 

For any spoofing attack to take place successfully, it is compulsory for the attacker to spoof four GPS channels at the receiver successfully and simultaneously. This is because the PVT solution solves the linear problem for four unknowns of X, Y, and Z coordinates plus time, and to do so, relies on the information it collects from at least four channels at the tracking block. According to Table \ref{table::datasets1}, SatGrid:G25 collected on Nov 8, 2019 at rooftop of \LocTwo{} includes data from eight PRNs. Hence we look into all possible combinations of these PRNs that could generate a successful PVT fix in the receiver. These combinations are called PRN-sets here, which include 70 cases for eight satellites.  

\begin{figure*}[thb!]
	\centering
	\begin{subfigure}[t]{.33\textwidth}
		\centering
		\includegraphics[height=1.75in,width=0.9\columnwidth]{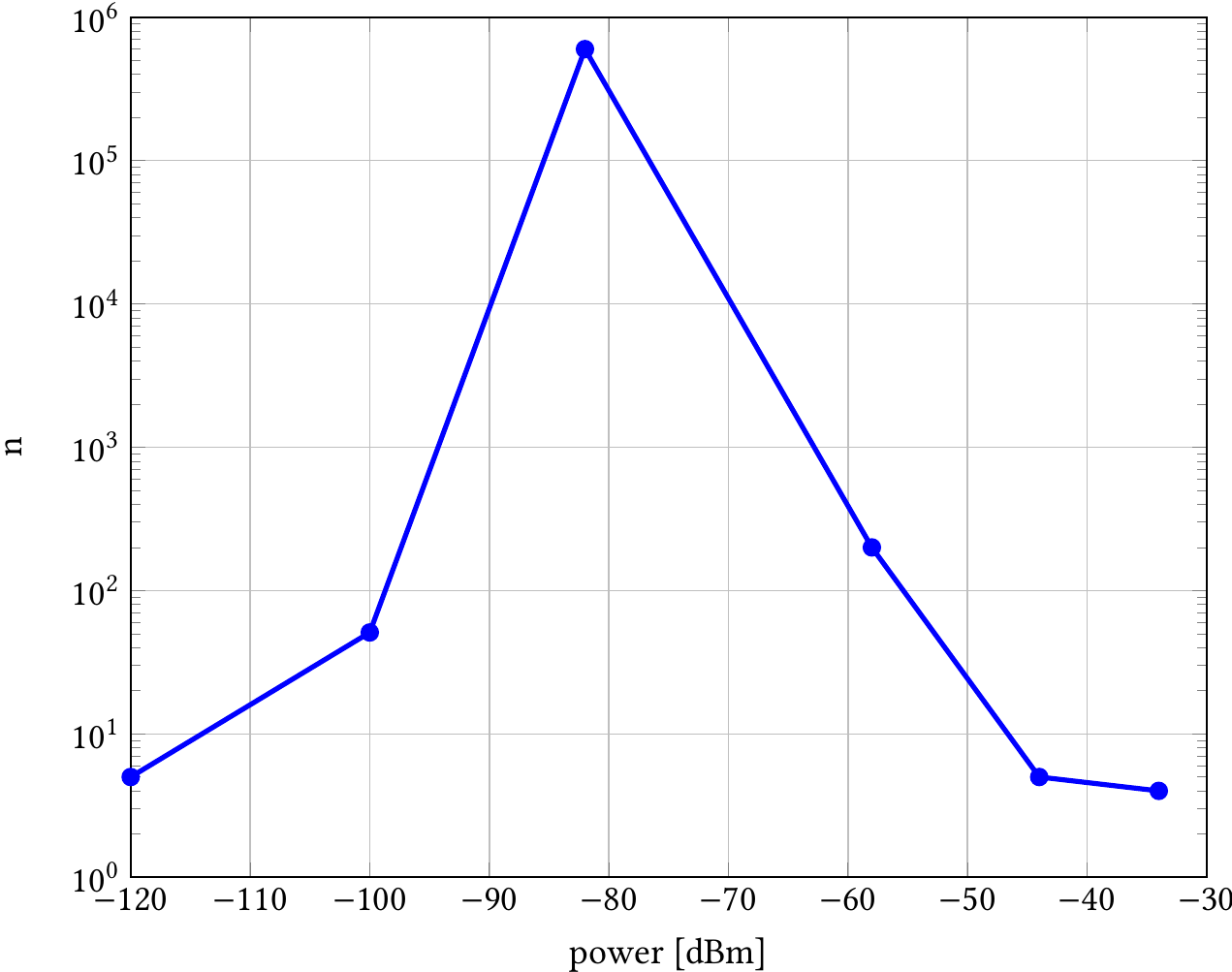}
		\caption{}
		\label{fig::zeroEER}
	\end{subfigure}
	\begin{subfigure}[t]{.33\textwidth}
		\centering
	\includegraphics[height=1.75in,width=0.9\columnwidth]{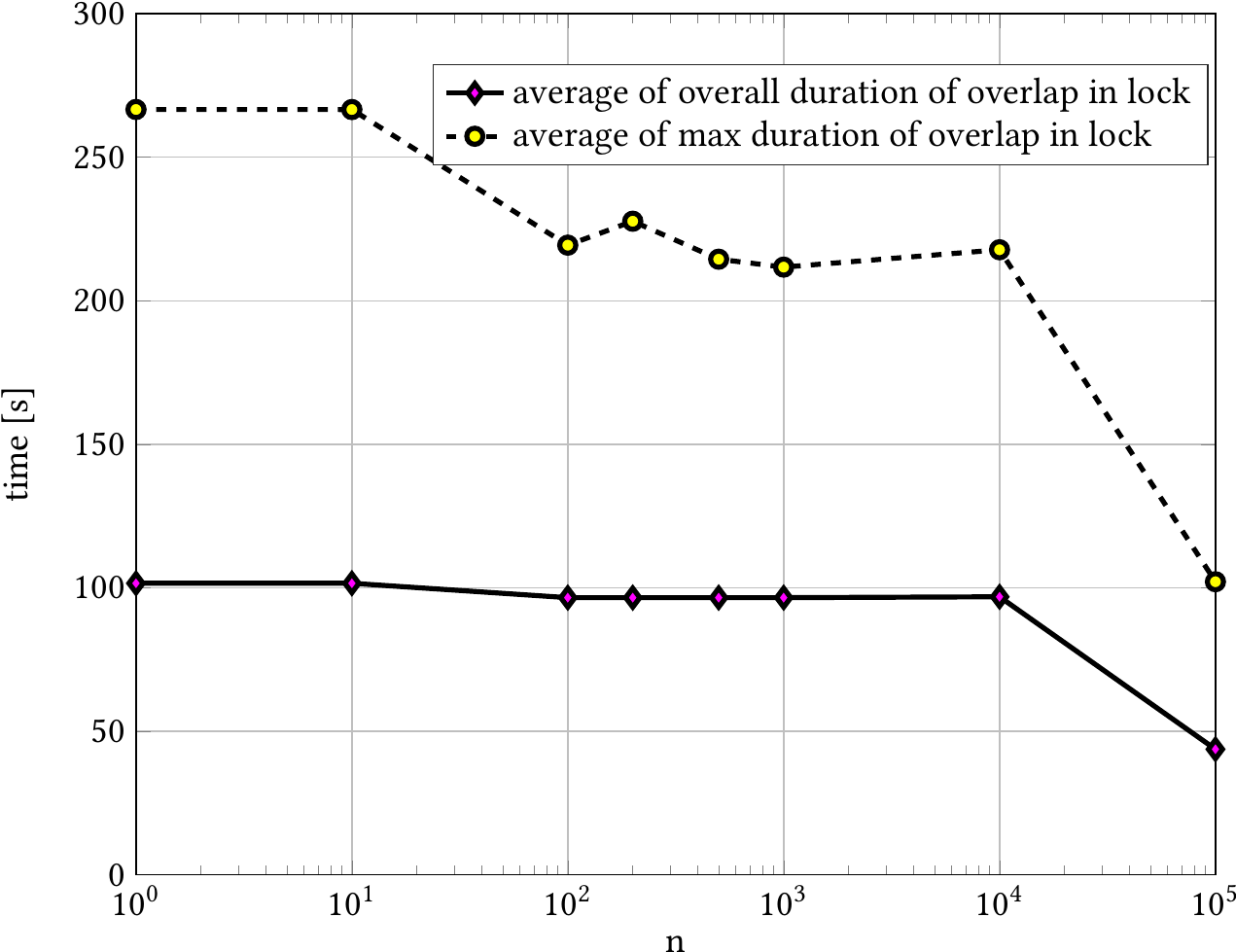}
		\caption{}
		\label{fig::times}
	\end{subfigure}
	\begin{subfigure}[t]{.33\textwidth}
		\centering
		\includegraphics[height=1.75in,width=0.9\columnwidth]{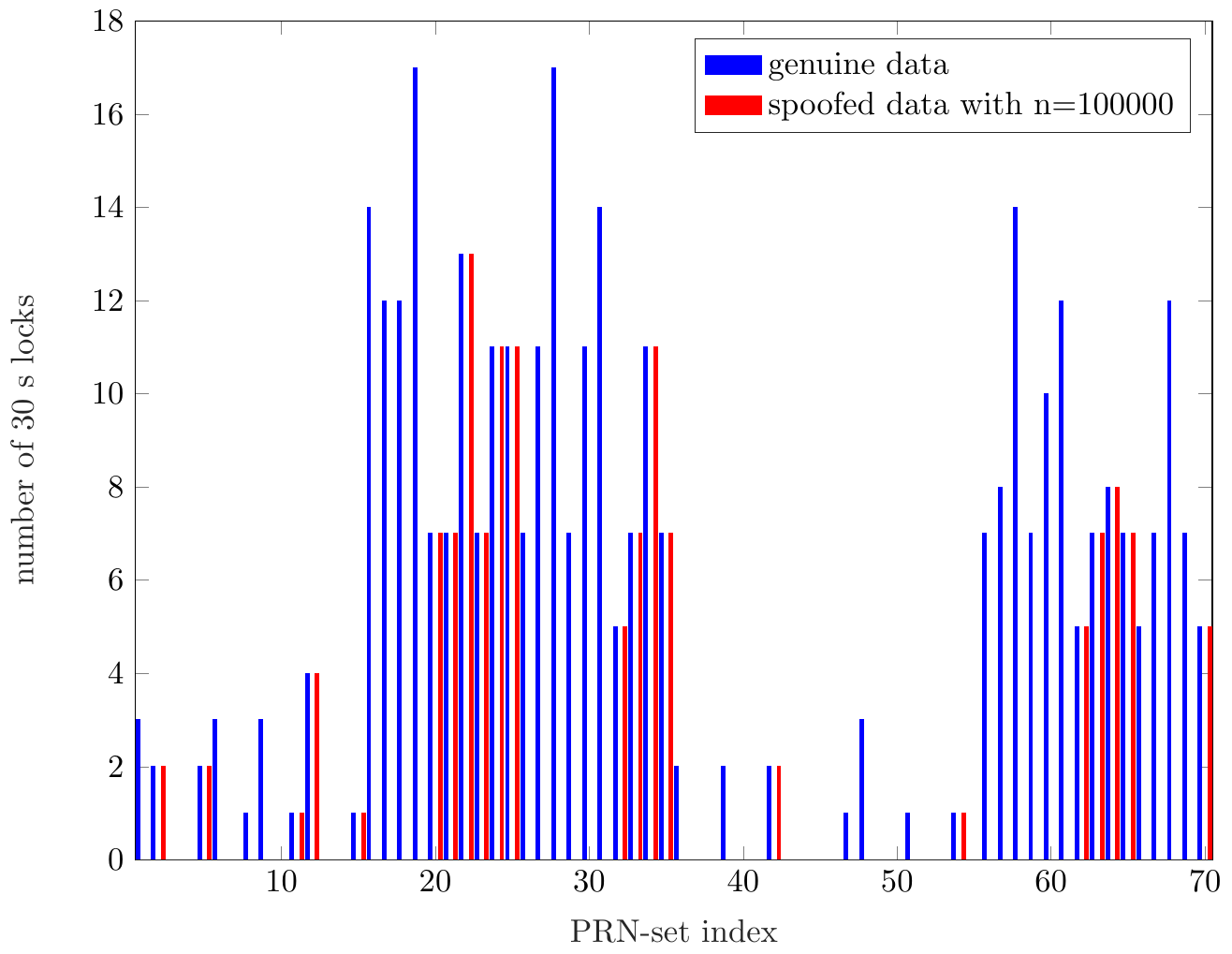}
		\caption{}
		\label{fig::nlock}
	\end{subfigure}
	
	\caption{(a) The number of required sample point observations (n) for Spotr to get EER of zero at different power levels of data collected on Sep 10, 2019 data at \LocTwo{}. (b)-(c) This analysis is performed on the matched-powered replay attack data, generated based off the genuine SatGrid:G25 Nov 8, 2019 data collected from the rooftop of \LocTwo{}.}\label{fig::timing}
\end{figure*}

Fig.~\ref{fig::times} shows the average of "maximum continuous spoofing time" for all the PRN-sets when Spotr exploits multiple sample points in order to perform the detection/tracking process. The X-axis shows the number of these sample points denoted by n. The average of "maximum continuous spoofing time" in this case reduces from \SI{100}{s} to \SI{47.3}{s} if 100,000 sample points are used by Spotr. 
The figure also reports the ``overall continuous spoofing time'' for all of the PRN-sets in the same graph, which is also reduced from \SI{266.6}{s} to \SI{101.2}{s} if n=100,000. This shows Spotr's ability to detect spoofing activities in \SI{47.3}{s} in presence of the strongest attacker. 

The navigation message consists of \SI{30}{s} frames which are 1,500 bits long. That is why the number of \SI{30}{s} locks that a spoofer is able to remain undetected by Spotr is a more accurate metric for performance evaluation. Fig.~\ref{fig::nlock} shows the number of \SI{30}{s} locks that the spoofer is able to generate for each PRN-set. Spotr is able to reduce the number of undetected locks from 360 occurrences when n=1, down to 131 when n=100,000. The number of locks for the genuine data is given as a bench mark here (plotted in blue bars) where there are 361 locks for all the PRN-sets. This gives a high level understanding of the performance of Spotr and does not translate directly to the continuous spoofing capabilities of the attacker. The attacker would be detected within the interval of \SI{30}{s} locks shown in Fig.~\ref{fig::nlock}.  

So far, we can conclude that the number of sample points to be averaged by Spotr, n,  is a critical parameter that directly influences the detection/tracking performance if the attacker is capable of adjusting its power levels. Fig.~\ref{fig::zeroEER} shows how this number changes with the SatGrid replay attacker's power level, when the EER holds the ideal value of zero for all the power levels. It increases as the under-powered attacker increases its strength to the matched-powered level with the genuine data, and reduces again as the power of the attacker gets far more than that of the genuine signals.  

\section{conclusion}\label{sec::conc}
In this work, we introduced a physical-layer identification based  spoofing detector for GPS satellites. Even though the spoofed signals are as closely phase aligned as possible with their authentic counterparts, they are never exactly the same as the genuine GPS signals. In fact a perfect carrier-phase alignment is impossible for a spoofer. We take advantage of anomalies that are introduced into the complex correlator outputs of a standard GPS receiver during a spoofing attack, without depending on the digital information or the GPS receiver's solution for position, velocity and time. Hence, our algorithm detects if spoofing attacks launched on civil GPS signals, and prevents the receiver from locking to the spoofed GPS signals in as few as \unit[47.3]{s} in worst case scenario. We validated our method by testing it on the de-facto standard of a publicly available repository of GPS signal spoofing traces called the Texas Spoofing Battery (TexBat), as well as our data (SatGrid) collected over several months at multiple locations in United States. More specifically, we are able to detect spoofing activities and track genuine signals over different times and locations and propagation effects related to environmental conditions.

\begin{acks}
This work was supported in part through funding from NSF IUCRC S2ERC (Award Number 1650540) and CACI International Inc. The authors wish to thank Dr.\ Ramin Baseri, Mr.\ John Lee, and Mr.\ Andrew Lee of CACI and Drs.\ Jeffrey Reed and Robert W. McGwier of Virginia Tech for their guidance on, and support of, the work.
\end{acks}

\bibliographystyle{ACM-Reference-Format}
\bibliography{ref}
\appendix 
%% list of datasets 
\section{appendix}
Table \ref{table::datasets1} details the authentic satellite data collection, and Table \ref{table::datasets2} the spoofed data collection for Sec.\ \ref{sec::data}.
% Tables \ref{table::datasets1} and \ref{table::datasets2} detail the genuine and spoofed data for Sec.\ \ref{sec::data}.
\begin{table*}[t]
\caption{Genuine datasets. SatGrid is the data that we collected at \LocOne{} and \LocTwo{}, and TexBat is the data provided by Radionavigation Lab in UT Austin detailed in Sec. \ref{sec::TexBat} collected by their 2015 hardware platform (D2). }\label{table::datasets1}
\small
\begin{tabular}{c|c|c|c|c|c|c}
\hline
\textbf{Dataset}        & \textbf{collection date} & \textbf{ multipath } & \textbf{duration} & \textbf{start time} & \textbf{location} & PRNs\\ \hline \hline

\textbf{TexBat:clean}     & July 2015               & No      & 420 s   & --     & Texas  &\{3,6,7,13,16,19,23\}         \\ \hline
\textbf{TexBat:clean}     & July 2015               & Yes     & 420 s   & --     & Texas  &\{9,15,18,22\}       \\ \hline

\textbf{SatGrid:G1}     & Sep 24, 2018                & No      & 60 min   & 8:11 am   & \LocOne{}   &\{1,7,8,9,11,18,27,28\}           \\ \hline
\textbf{SatGrid:G2}     & Sep 25, 2018                & No      & 60 min   & --    & \LocOne{}   &\{1,7,8,9,11,18,27,28\}          \\ \hline
\textbf{SatGrid:G3}     & Sep 26, 2018                & No      & 60 min   & --    & \LocOne{}   &\{1,7,8,9,11,18,27,28\}           \\ \hline
\textbf{SatGrid:G4}     & Sep 28, 2018                & No      & 60 min   & 4:10 pm   & \LocOne{}   &\{1,7,8,9,11,18,27,28\}           \\ \hline

\textbf{SatGrid:G5}     & Dec 15, 2018                & No      & 60 min  & 12 pm     & Missouri      &\{10,14,20,21,32\}       \\ \hline
\textbf{SatGrid:G6}     & Dec 15, 2018                & No      & 60 min  & 2 pm      & Missouri      &\{14,22,25,31,32\}       \\ \hline
\textbf{SatGrid:G7}     & Dec 15, 2018                & No      & 60 min  & 8 pm      & Missouri      &\{7,8,9,27,30\}       \\ \hline

\textbf{SatGrid:G8}     & Dec 16, 2018                & No         & 60 min   & 2 am     & Missouri   &\{2,6,12,17,19\}          \\ \hline
\textbf{SatGrid:G9}     & Dec 16, 2018                & No         & 60 min   & 3:30 pm  & Missouri   &\{3,14,22,26,31\}          \\ \hline
\textbf{SatGrid:G10}     & Dec 16, 2018                & No         & 60 min   & 7 pm     & Missouri    &\{7,8,9,23,27\}         \\ \hline
\textbf{SatGrid:G11}     & Dec 16, 2018                & No         & 60 min   & 9 pm     & Missouri    &\{1,7,8,11,30\}         \\ \hline
\textbf{SatGrid:G12}     & Dec 16, 2018                & No         & 60 min   & 11 pm    & Missouri   &\{1,17,19,28,30\}          \\ \hline

\textbf{SatGrid:G13}     & Dec 17, 2018                & No      & 60 min     & 6 am   & Missouri   &\{7,11,17,28,30\}          \\ \hline
\textbf{SatGrid:G14}     & Dec 17, 2018                & No      & 60 min     & 12 pm  & Missouri   &\{10,14,20,31,32 \}          \\ \hline
\textbf{SatGrid:G15}     & Dec 17, 2018                & No      & 60 min     & 10 pm  & Missouri   &\{2,5,13,15,29\}          \\ \hline

\textbf{SatGrid:G16}     & Dec 18, 2018                & No      & 60 min   & 4 am    & Missouri   &\{2,5,6,12,25\}          \\ \hline
\textbf{SatGrid:G17}     & Dec 18, 2018                & No      & 60 min   & 11 am   & Missouri   &\{10,14,18,20,32\}          \\ \hline
\textbf{SatGrid:G18}     & Dec 18, 2018                & No      & 60 min   & 2 pm    & Missouri   &\{3,14,22,31,32\}          \\ \hline
\textbf{SatGrid:G19}     & Dec 18, 2018                & No      & 60 min   & 11 pm   & Missouri   &\{1,11,17,19,28\}          \\ \hline

\textbf{SatGrid:G20}     & Dec 19, 2018                & No     & 60 min   & 9 am   & Missouri    &\{10,15,20,21,24\}         \\ \hline
\textbf{SatGrid:G21}     & Dec 20, 2018                & No     & 60 min   & 9 pm   & Missouri    &\{1,11,13,18,28\}         \\ \hline

\textbf{SatGrid:G22}     & Aug 23, 2019                & yes     & 10 min   & 4 pm     & \LocTwo{} (urban)     & \{2,13,15,21,29\}       \\ \hline
\textbf{SatGrid:G23}     & Sep 10, 2019                & No      & 45 min   & 3 pm     & \LocTwo{} (football field) & \{2,5,8,15,17,21,24,29\}    \\ \hline
\textbf{SatGrid:G24 \footnotemark{}  }     & Nov 8, 2019                  & No      & 45 min   & 10 am    & \LocTwo{} (rooftop)     & \{2,24,13,15,20,5,21,29\}         \\ \hline
\textbf{SatGrid:G25}     & Nov 8, 2019                  & No      & 50 min   & 11 am    & \LocTwo{} (rooftop)     & \{10,24,13,15,20,5,21,29\}         \\ \hline

\hline
\end{tabular}

%\tablefootnote{}
\end{table*}
\begin{table*}[t]
\caption{Spoofed datasets. There are different hardware platforms (fingerprinters) that are used for generating them. The fingerprinter of TexBat on 2012 (D1) is different from TexBat 2015 (D2). SatGrid attacker also has a separate hardware (D3).}\label{table::datasets2}

\small
\begin{tabular}{c|c|c|c|c|c|c|c}
\hline
\textbf{Dataset}        & \textbf{collection date} & \textbf{spoofing type}& \textbf{threat model} & \textbf{spoofing power status} & \textbf{ multipath } & \textbf{duration}  & \textbf{location} \\ \hline \hline

\textbf{TexBat:S1}     & Sep 2012                 & both   & Replay                  & under-powered                    & No    & 420 s         & Texas            \\ \hline
\textbf{TexBat:S2}     & Sep 2012                 & time   & Replay                & over-powered           & No    & 420 s         & Texas            \\ \hline
\textbf{TexBat:S3}     & Sep 2012                 & time   & Replay               & matched-powered        & No    & 420 s         & Texas            \\ \hline
\textbf{TexBat:S4}     & Sep 2012                 & position & Replay              & matched-powered        & No    & 420 s         & Texas            \\ \hline
\textbf{TexBat:S5}     & Sep 2012                 & time     & Replay              & over-powered           & Yes   & 420 s         & Texas            \\ \hline
\textbf{TexBat:S6}     & Sep 2012                 & position & Replay              & matched-powered        & Yes   & 420 s         & Texas            \\ \hline
\textbf{TexBat:S7}     & July 2015                & time     & Replay              & matched-powered        & No    & 420 s         & Texas            \\ \hline
\textbf{TexBat:S8}     & July 2015                & time     & Replay              & matched-powered        & No    & 420 s         & Texas            \\ \hline

\textbf{SatGrid:S1} & Sep 24, 2018             & both      & Spoofing              & over-powered              & No     & 60 min      & \LocOne{}               \\ \hline
\textbf{SatGrid:S2} & Sep 25, 2018             & both      & Spoofing              & over-powered              & No     & 60 min      & \LocOne{}               \\ \hline
\textbf{SatGrid:S3} & Sep 26, 2018             & both      & Spoofing              & over-powered              & No     & 60 min      & \LocOne{}              \\ \hline
\textbf{SatGrid:S4} & Sep 28, 2018             & both      & Spoofing              & over-powered              & No     & 60 min      & \LocOne{}              \\ \hline

\textbf{SatGrid:S5} & Dec 15, 2018             & both     & Spoofing               & over-powered         & No      & 60 min      & Missouri               \\ \hline
\textbf{SatGrid:S6} & Dec 16, 2018             & both     & Spoofing               & over-powered         & No      & 60 min     & Missouri               \\ \hline

\textbf{SatGrid:S7}   & Aug 23, 2019                & both   & Replay           & adjusted-power          & No     & 10 min   & \LocTwo{}              \\ \hline
\textbf{SatGrid:S8}   & Sep 10, 2019                & both   & Replay           & adjusted-power          & No     & 45 min   & \LocTwo{}              \\ \hline

\textbf{SatGrid:S9}   & Nov 8, 2019                & both    & Replay          & adjusted-power           & No     & 50 min   & \LocTwo{}              \\ \hline
\textbf{SatGrid:S10}   & Nov 8, 2019                & both    & Replay          & adjusted-power           & No     & 50 min   & \LocTwo{}              \\ \hline

\end{tabular}
\end{table*}
\newpage
\footnotetext{SatGrid:G24 and SatGrid:G25 have high fidelity timestamps.}

\end{document}